\documentstyle[eqsecnum,prb,aps,epsf]{revtex}

\begin{document}

\twocolumn

\def \der{\partial}
\def \T{\mathop{{\rm T}_\tau}\nolimits}
\def \tr{\mathop{\rm tr}\nolimits}
\def \Torb{T_K^{\rm orb}}
\def \tens{\underline}
\def \skip{\vskip 0.5truecm}
\def \k{{\bf k}}
\def \v{{\tens v}}
\def \V{{\tens V}}
\def \U{{\tens U}}
\def \sig{{\tens \sigma}}
\def \t{{\tens t}}
\def \rom{{\tens \varrho}}
\def \M{{\tens M}}
\def \q{{\bf q}}
\def \0{{\tens 0}}
\def \dalpha{{\delta\alpha}}
\def \dv{{\tens {\delta v}}}
\def \r{\varrho}
\def \etam{{\tens \eta}}
\def \dag{{+}}
\def \L{{\tens L}}
\def \nn{{\nonumber}}

\draft
\preprint{}
\title{Solution of the Multichannel Coqblin-Schrieffer Impurity Model and
Application to Multi-Level systems}

\author{Andr\'es Jerez\cite{ill}}
\address{University of Oxford, Department of Physics, Theoretical
Physics, 1 Keble Road, \\  Oxford OX1 3NP, United Kingdom}
\author{Natan Andrei}
\address{Department of Physics and Astronomy, Rutgers University,
Piscataway, NJ 08855, USA}
\author{Gergely Zar\'{a}nd}
\address{Institute of Physics, Technical University of Budapest, \\
H 1521 Budafoki \'{u}t 8., Budapest, Hungary}

\maketitle

\begin{abstract}
A complete Bethe Ansatz solution of the $SU(N)\times SU(f)$ Coqblin
Schrieffer model and a detailed analysis of some physical applications of the
model are given. 
As in the usual multichannel 
Kondo model a variety of Fermi liquid and non-Fermi liquid (NFL) fixed points 
is found, whose nature depends on the impurity representation, $\mu$.
For $\mu=f$ we find  a Fermi liquid fixed 
point, with the impurity spin  completely screened. For $f>\mu$ the 
impurity is overscreened and the model has NFL properties. The form the NFL
behavior takes depends on the $N$ and $f$:
 for $N\le f$ the specific heat and the susceptibility are dominated by
the NFL contributions,  
 for $N>f$ the leading contributions are Fermi-liquid like and the 
NFL behavior can be seen only to subleading order, while for $N=f$ the
behavior is marginal.
We also analyze  the possibility of physical realizations.
We show by a detailed renormalization group  and $1/f$ analysis that 
the tunneling $N$-state problem can be mapped into the $SU(N)\times SU(f)$
exchange model, and discuss the subtle differences between the two models.
As another physical realization we suggest a double quantum dot structure
that can be described by means of an $SU(3)\times SU(2)$ model if the 
parameters of the dots are tuned appropriately.  
\end{abstract}
\pacs{75.20.Hr, 75.30.Mb, 71.10.Hf}



\tableofcontents

\section[Introduction]{Introduction}

The multichannel Kondo model \cite{Nozieres} is the simplest
impurity model with non-Fermi liquid behavior. Originally introduced
to describe "real metals" with magnetic impurities, 
its applications go beyond the study of dilute
magnetic alloys. For instance,
it has been known for some time that systems consisting of heavy
atoms tunneling between two neighboring sites and interacting
with conduction electrons are a realization of the two-channel Kondo
model \cite{ZawVlad}. Another realization is the quadrupolar
Kondo effect in context of heavy fermions \cite{cox1}. A detailed account of
various aspects and applications of the multichannel Kondo model is given in
Ref.~\onlinecite{coxzawa}.

For materials such as $Pb_{1-x}Ge_xTe$ or $K_{1-x}Li_x Cl$
alloys, the tunneling may occur between an arbitrary number of levels.
 Such systems could be modeled using a multichannel
 version of the
Coqblin-Schrieffer model, a $SU(N)\times SU(f)$ Kondo
model\cite{ZarPRL}. Here, $N$ is the number of spin degrees of freedom, and $f$
is the number of channels, or $flavor$ degrees of freedom.

In this article we present an exact solution of the $SU(N) \times
SU(f)$ Kondo model, and study the thermodynamic properties of the
system. We obtain the leading exponents for the
impurity contribution to the magnetic susceptibility and specific heat
for arbitrary 
$N$ and $f$. We also
discuss the effects of channel anisotropy, which might drive the
system from a fixed point with $\gamma \equiv f/N > 1$ to a new fixed
point where
$\gamma^{eff} < 1$.

\section[The Model]{The Model}

The Multichannel Coqblin-Schrieffer (MCCS) model describes
 electrons carrying two sets of internal degrees of freedom,
to be denoted spin and flavor (or channel number), interacting
 with an impurity carrying only spin. The impurity is localized
at a point chosen to be the
origin. The Hamiltonian reads,

\begin{eqnarray}
{\cal H} &=& -i \sum_{a}^{N} \sum_{m}^{f}
\int_{-\infty}^{\infty} 
\psi^{\dagger}_{a,m}(x)\partial_x \psi_{a,m}(x)\; dx  \nonumber 
\\ &+& 2 \sum_m^{f} J_m 
\sum_{\alpha}^{N^2-1} \sum_{a,b}^{N}
\psi_{a,m}^{\dagger}(0)\left(T_{\alpha}^{(\Box)}\right)_{a,b}
\psi_{b,m}(0)   \nonumber \\ &\times &
\sum_{a',b'}^{dim(\mu)} 
\chi_{a'}^{\dagger}\left(T_{\alpha}^{(\mu)}\right)_{a',b'}\chi_{b'}.
\label{ham}
\end{eqnarray}
Both $\psi_{a,m}^{\dagger}(x)$ and $\chi_{a}^{\dagger}$
are fermionic fields,  the former creates an electron at $x$ with 
spin index
$a$ and flavor index $m$, while  the latter creates the 
impurity at $x=0$. Imposing the condition $ \sum_a \chi_{a}^{\dagger}\chi_{a}=1$,
we have that
$\chi_{a'}^{\dagger}\left(T_{\alpha}^{(\mu)}
\right)_{a',b'}\chi_{b'}$ represents the impurity spin 
operator in a
representation of $SU(N)$ specified by a particular
choice of the matrices $T_{\alpha}^{(rep)}$, where 
 the index $\alpha$ runs from 1 to
$N^2-1$, the number of generators of $SU(N)$.
 We will restrict
ourselves to the case in which the electrons are in the fundamental
representation (denoted by $(\Box)$), and the impurity is in the
totally symmetric representation obtained from the direct product of 
$\mu$ fundamental representations. 

The physical realizations discussed in the present paper
 correspond to the simplest
case, $\mu=1$; the investigation of the $\mu>1$ cases gives us important 
insight into the general structure of the model and allows, in particular, for a comparison
with the results obtained for the multichannel Kondo model
with impurity spin $S>1/2$.

 In most of
this paper we will study the isotropic model, $J_m=J$, with the symmetry
$U(1)^{charge}\times SU(N)^{spin} \times SU(f)^{flavor}$.
 We will also assume that the different flavor levels are
equally populated, $N^e=fN_0$.

In what follows shall solve the complete model and among other things,
study its low-energy physics. As is well known, the low energy behavior
 of a system can often be described in terms of effective hamiltonians,
 that are simpler
than the starting hamiltonian; these are usually referred to as fixed points. We
shall determine their properties from the exact solution. 
We shall find that the model possesses a variety of fixed points 
(or low energy regimes),
 whose nature depends on 
the symmetry structure in the flavor sector and on the spin representation 
($\mu$), generalizing
the familiar  $N=2$ case (the multichannel Kondo model
\cite{Nozieres,adkondo,tw1,AL}. As previously, we shall identify the mechanism underlying
 the appearance of these
fixed points as {\it dynamical fusion} by which electrons form 
 spin complexes whose
  interaction with the impurity leads to a new behavior in the infrared
\cite{adkondo}. Each complex consists of $f$ electrons fused 
into a local objects that transforms
 according to one-row Young Tableaux of length $f$.

Within the Bethe-Ansatz approach  a precise description of the formation 
of these composites can be given. 
 The  linearized hamiltonian propagates separately
the  charge-spin-flavor degrees of freedom  that make up the electron.
Therefore the effect of flavor on the spin degrees of freedom is recovered only
in the physical space. To follow the dynamic 
coupling of spin and flavor  we add some curvature which maintains the identity
of the electron while allowing its components to interact. It has the form 
$H_{\Lambda} = \frac{1}{2 \Lambda} \sum_{a} \sum_{m}
\int_{-\infty}^{\infty}  \psi_{a,m}^\dagger (x) \partial_x^2
\psi_{a,m} (x)dx $, where $\Lambda$ is the curvature scale which is sent to 
infinity at the end of the calculation. Adding this term allows for the formation 
of bound states in the flavor singlet channel, which interact  strongly with the impurity,
and determine the low-energy dynamics even after the curvature is removed. A close analogy
is  a small magnetic field introduced to probe for magnetization, which may survive 
after the field is removed. Imposing a cut-off $D$ on the momentum variables
 guarantees the finiteness of the energy.
Other terms need to be added to the hamiltonian to maintain integrability, terms which we shall
see below are irrelevant.

 Already for free 
fields the resulting theory is quite involved, and even the counting of
states is not trivial\cite{destri}. Nevertheless, the charge-spin-flavor separated 
basis is the natural one for the non-interacting problem, as we shall see later:
 it is the form to which the eigenstates tend when the interaction is turned off.
We thus introduce the following elements:

\begin{itemize}
\item
A second derivative term with a curvature scale, $\Lambda$, 
\begin{eqnarray} 
H_{\Lambda} = \frac{1}{2 \Lambda} \sum_{a} \sum_{m}
\int_{-\infty}^{\infty}  \psi_{a,m}^\dagger (x) \partial_x^2
\psi_{a,m} (x)dx 
\label{second} 
\end{eqnarray}
which breaks charge-spin-flavor (CSF) separation of the linear spectrum.
 Once the electron composites
are formed, and the low-energy spectrum of the theory is identified,
the scale is taken to infinity.
 
Adding the term (\ref{second}) also imposes
 restrictions on the form of the eigenstates which can be expressed 
in terms of the following counterterms without which 
 the model is not integrable for finite $\Lambda$:

\item
An electron-electron interaction term, of the form
\begin{eqnarray}
\!\!\!\!\!\!\!\!
2\tilde{J}\sum_{m,m'} \sum_{a,a'} & &
\int_{-\infty}^{\infty} \!\!  
\psi_{a,m}^{\dagger} (x) \psi_{a',m'}^{\dagger} (x) \psi_{a,m'} (x)
\psi_{a',m} (x)dx  \nonumber \\ & &
\label{jt}  
\end{eqnarray}

When no impurity is present $\tilde{J}$ can be chosen arbitrarily since
the term
has no effect on the linear spectrum. The
linearized spectrum has a large degeneracy, and the
inclusion of (\ref{second}) and (\ref{jt}) will provide a way to find the
eigenstates.

\item
A counterterm $H_{cc}$, of the form 
\begin{eqnarray}
H_{cc} &=& - \frac{1}{\Lambda}\sum_{m}^f \sum_{a}^{N} 
\int_{-\infty }^\infty \!\!\!\!  \psi _{a,m}^{\dagger
}(x) V(x)\psi
_{a,m}(x)dx 
\label{hcc}  
\end{eqnarray}
with
\begin{eqnarray}
V(x)=\frac x{|x|}(\delta ^{\prime }(x^{+0})+\delta ^{\prime }(x^{-0})),
\label{vx}
\end{eqnarray}
needs to be  added to the Hamiltonian in order to preserve
integrability at the origin;
this term vanishes 
once the curvature is removed, and plays no further role in the problem. 
\end{itemize}

\subsection[First Quantized Hamiltonian]{First Quantized Hamiltonian}

A general Fock state of $N^e$ electrons and one impurity
 can be written in the following form:
\begin{eqnarray*}
|F> &=& \sum_{\{m_j\}} \sum_{\{a_j\},b} \int_{-\infty}^{\infty}
(\prod_{j} dx_j ) F_{\{a_j\},b}^{\{m_j\}}(\{x_j\})
\chi_b^{\dagger}(0) \\ &\times & 
\prod_{j=1}^{N_e} \psi^{\dagger}_{a_j,m_j}(x_j) |0>.
\end{eqnarray*}
In order for it to be an eigenstate the amplitudes $F$ must 
satisfy the 
equation $hF=EF$, where
the differential operator $h$, known as the {\it first quantized form
of the Hamiltonian}, takes the form 
\begin{eqnarray*}
h &=& \sum_{j=1}^{N^e} \{-i \partial_j + \frac{1}{2\Lambda} \partial_j^2 
+ 2J\delta(x_j)
\sum_{\alpha}^{N^2-1}
\left(T_{\alpha,j}^{(\Box)}\right)\left(T_{\alpha}^{(\mu)}\right) \}
\\  &+& \sum_{l<j} 2 \tilde{J} \delta(x_l-x_j)
(P_{lj}-{\cal P}_{jl}) - \sum_{j=1}^{N^e} \frac{1}{\Lambda} V(x_j),
\end{eqnarray*}
with $P_{jl}$(${\cal P}_{jl}$)  the spin(flavor) exchange operator,
\begin{eqnarray*}
P_{ab,cd} &=& \delta_{ad}  \delta_{bc}, \\
{\cal P}_{m_1 m_2,m_3 m_4} &=& \delta_{m_1 m_4} \delta_{m_2 m_3}  .
\end{eqnarray*} 
The fundamental representation $(\Box)$ is carried by the
electron $j$ and the $(\mu)$ representation by the impurity. When the
latter
is also in the fundamental representation i.e. $\mu =1$ the
hamiltonian can be rewritten as,
\begin{eqnarray*}
h &=& \sum_{j=1}^{N^e} -i \partial_j + (\Lambda^{-1}) \partial_j^2 
+ 2J \delta(x_j) P_{j0} 
\\ &+& \sum_{l<j} 2 \tilde{J} \delta(x_l-x_j)
(P_{lj}-{\cal P}_{jl}) +  \sum_{j=1}^{N^e} \frac{1}{\Lambda} V(x_j).
\end{eqnarray*}

\subsection[S-matrices]{S-matrices}

We will assume for now that both the electrons and the
impurity are in the fundamental representation of $SU(N)$. The
eigenstate amplitudes are combinations
of plane waves with {\it pseudo-momenta} $k_j$, ($j=1,...,N^e$), and
have coefficients that depend on the ordering of the electrons, and on
the spin and the flavor indices. These coefficients are related through
products of electron-impurity and electron-electron S-matrices that we
will derive now. Consider first  the wavefunction describing 
one electron (denote it by $j$) interacting with the impurity (denote
it by $0$),
\begin{eqnarray} 
F_{a_j,a_0}^{m_j}(x_j) = e^{ik_jx_j}\left(A_{a_j,a_0}^{m_j}\theta(-x_j)+
B_{a_j,a_0}^{m_j} \theta(x_j)\right). 
\label{amp1}
\end{eqnarray} 
 Applying  $h$ to it we have (we drop the indices in the amplitudes)
\begin{eqnarray}
hF(x_j) &=& (k_j-\frac{k_j^2}{2\Lambda})F(x_j) \label{amp2} \\ 
& & (-i(1-\frac{k_j}{\Lambda}) (B-A) + J P_{j0}(B+A) )\delta(x_j) 
\nonumber \\ && -\frac{1}{2\Lambda}(B-A)\delta'(x_j) e^{i k_j x_j} +
\frac{1}{\Lambda} V(x_j) F(x_j). \nonumber
\end{eqnarray}
 $F$ is  an eigenstate of $h$, with eigenvalue $E_j=
k_j(1- \frac{k_j}{2\Lambda})$, if the terms in the second and third
lines in (\ref{amp2}) vanish. The last two terms cancel each other
due to the form of (\ref{vx}). The terms in the second line of
(\ref{amp2}) cancel if the amplitudes $A$ and $B$ are related by
 the electron-impurity S-matrix  $B=S_{j0} A$, where 
 $S_{j0} = \left(S_{j0}\right)_{a_j,a_0}^{a_j',a_0'}$ is given by
\begin{eqnarray}
S_{j0} &=& \frac{i(1-k_j/\Lambda)+ J P_{j0}}
{i(1-k_j/\Lambda)- J P_{j0}} \label{si} \\ &=&
\left(\frac{i(1-k_j/\Lambda) + J} 
{i(1-k_j/\Lambda) -J}\right)\left( \frac{i(1-k_j/\Lambda) + J P_{j0}}
{i(1-k_j/\Lambda) + J}\right)^2. \nonumber
\end{eqnarray}

Defining
\begin{eqnarray*}
c \equiv \frac{2J}{1-J^2}, ~~~~ g(x) \equiv \frac{1-x}{1-J^2} \left( 1
- \frac{J^2}{(1-x)^2} \right),
\end{eqnarray*}
we can write
\begin{eqnarray*}
S_{j0} = e^{-i \arctan \frac{c}{g(k_j/\Lambda)}} \left(
\frac{g(k_j/\Lambda)-icP_{j0}}{g(k_j/\Lambda)-ic}\right),
\end{eqnarray*}
(notice that $\arctan c = 2\arctan J$). Eventually we will send the
cutoff to infinity. Therefore,  expanding $g(k/\Lambda)$ to first
order in $1/\Lambda$
\begin{eqnarray*}
g(k/\Lambda) \sim 1 - \left(\frac{1+J^2}{1-J^2} \right) \frac{k}{\Lambda},
\end{eqnarray*}
we have
\begin{eqnarray}
S_{j0} \sim e^{i \arctan \frac{c}{1+\lambda_j}}
\left(\frac{\lambda_j-1+ic P_{j0}}{\lambda_j-1+ic}\right), 
\label{eis} 
\end{eqnarray}
where
\begin{eqnarray*}
\lambda_j = \left( \frac{1+J^2}{1-J^2} \right) \frac{k_j}{\Lambda},
\end{eqnarray*}
In the scaling limit, $J$ and $c$ have the same scaling behavior.

We now consider the case of two electrons. We generalize the procedure
followed in the case on one electron: divide the 
configuration space into regions inside each of which
there is no interaction and the wavefunction is a superpositions of 
plane waves.
They are six such regions
in this case, corresponding
to the ordering of three objects: two electrons and an impurity,
and we label them by permutations $Q \in S_3$.
For example, the element $Q= (1,0,2)$ labels the region where electron 1 is to
 the left of the impurity and electron 2 is to its right. We also introduce
 the notation
$\theta (x_Q)$ to denote a function that takes the value 1 in the region $Q$
 and zero elsewhere.

The two electron wave function is then of the form (Bethe-Ansatz),
\begin{eqnarray*}
F^m_a(x)={\cal A} e^{i(k_1x_1+k_2x_2)}\sum_Q \theta (x_Q) A^Q_{a,m},
\end{eqnarray*}
where $m = (m_1,m_2)$ and $a= (a_1,a_2,a_0)$ and ${\cal A} $ is the antisymmetrizer.
The amplitudes in the various regions are connected by $S$-matrices,
 e.g. $S^{01} A^{012}= A^{102}$, where $S^{01}$, the electron-
impurity $S$-matrix has been already determined in the one electron problem.
For this Ansatz to be consistent it must satisfy  the Yang-Baxter Relations,
\begin{equation}
S^{ij}S^{i0}S^{j0}=S^{j0}S^{i0}S^{ij} .
\label{ybe}
\end{equation}
guaranteeing that the two paths from $(1,2,0)$ to $(0,2,1)$ yield the 
same answer.
 
What is the electron-electron $S$-matrix, $ S^{ij}$?
There is no direct electron-electron
interaction
term in the Hamiltonian (\ref{ham}) , and
one may be tempted to adopt  the naive choice
 $S^{ij} =I$ for the
scattering
matrix of electrons $i$ and $j$. Nevertheless, electron
correlations are induced through the impurity. These show
 up immediately, since the naive choice does not
satisfy the Yang-Baxter Relations:  $S^{j0}$ and $S^{i0}$ do not commute. 
This non commutativity  captures some
important aspects of the model: after electron $i$ crossed the
impurity the latter is left in a different state than before. Hence the
state in which
electron
$j$  finds the impurity depends on whether it crosses the impurity
before or after electron $i$. Herein lies the difference between a
system of electrons interacting with a fixed potential (a one-body
problem since all electron see the same potential) and a Kondo system,
where the impurity correlates the motion of all electrons.

 Are we allowed
to introduce a scattering matrix $S^{ij}$ to satisfy the Yang-Baxter Relations?
We proceed now to show that this is indeed the case, namely, the introduction of
an electron-electron scattering matrix would not modify the original problem we set
out to solve.
 Consider first the space of free electrons with a linearized  Hamiltonian.
The space is highly degenerate: for example the energy $E=k_1+k_2$ in
the two electron space corresponds
to a wave function $F=\sum_q  e^{i(k_1+q)x_1 + i(k_2-q)x_2}A_q $ for any choice
of coefficients $A_q$. Equivalently, we can pick a basis of the form
$F=  e^{ik_1 x_1 +ik_2 x_2} \left( \theta(x_1-x_2)+ S^{12}
\theta(x_2-x_1)\right) A$. The choice of $S^{12}$ is
arbitrary in the two electron space, but if we wish
to proceed to construct three (and more) electron wavefunctions
then  the scattering 
matrices satisfy must satisfy the YBE for electrons, $S^{ij}S^{ik}S^{jk}=
S^{jk}S^{ik}S^{ij}$.
 When the Kondo interaction is turned on, the matrix 
$S^{j0}$ is fixed by the interaction, which in  turn picks
the electron basis through the Yang-Baxter Relations (\ref{ybe}).

 When the cut-off is present
 part of the degeneracy is removed already at the free electron level
but the procedure still goes through.
Consider the model for two electrons
away from the impurity.
\begin{eqnarray}
h &=& -i \partial_j- i\partial_l+ \frac{1}{2\Lambda} \partial_j^2 
+\frac{1}{2\Lambda} \partial_l^2 \nonumber \\ &+&
 2 \tilde{J} \delta(x_l-x_j)
(P_{lj}-{\cal P}_{lj}). 
\end{eqnarray}
This cut-off Hamiltonian  is in the same universality class as the free
linearized hamiltonian and possesses the same spectrum
when the cut-off is sent to infinity; its particular form was chosen so that the S-matrix
it defines does indeed satisfy the (\ref{ybe}).
Again, we divide configuration space into two regions:
\begin{eqnarray*}
\lefteqn{F_{\{a_j,a_l\}}^{\{m_j,m_l\}}(x_j,x_l) =} & & \\ & &
e^{i(k_jx_j +
k_lx_l)}\left(A_{\{a_j,a_l\}}^{\{m_j,m_l\}} \theta(x_l-x_j)+
B_{\{a_j,a_l\}}^{\{m_j,m_l\}} \theta(x_j-x_l)\right) \! \!, 
\end{eqnarray*}
and study the eigenvalue equation $hF=EF$. We have
\begin{eqnarray}
hF &=& \left( (k_j-\frac{k_j^2}{2\Lambda}) +
(k_l-\frac{k_l^2}{2\Lambda}) \right)F \label{amp3}
\\ && + (-i(B-A)-i(A-B) \nonumber \\ && 
+i(\frac{k_j}{\Lambda} - \frac{k_l}{\Lambda})(B-A)) \delta(x_j-x_l)
e^{i(k_j+k_l)x_j} \nonumber \\ && + \tilde{J} ( P_{jl}-{\cal P}_{jl})
(A+B) \delta(x_j-x_l) e^{i(k_j+k_l)x_j} \nonumber \\
& & + \frac{1}{2\Lambda} \left( (B-A)
+(A-B)\right)\delta'(x_j-x_l)  e^{i(k_jx_j+k_lx_l)}. \nonumber 
\end{eqnarray} 
The last line is identically zero; counterterms of the form
(\ref{hcc}) are only necessary when the particles involved have
different velocities. The rest of the terms proportional
to $\delta(x_j-x_l)$ cancel if the amplitudes in the different regions
are related by 
the following electron-electron S-matrix,
\begin{eqnarray*}
S_{jl} = \frac{i\alpha_{jl} +\tilde{J}( P_{jl}- {\cal P}_{jl})}
{i\alpha_{jl}-\tilde{J}( P_{jl} - {\cal P}_{jl})},
\end{eqnarray*}
where $\alpha_{jl} \equiv (k_l-k_j)/\Lambda$. Such a S-matrix can be
written as 
\begin{eqnarray*}
S_{jl} = \frac{\alpha_{jl}-2i\tilde{J} P_{jl}}{\alpha_{jl}-2i\tilde{J}}
\frac{\alpha_{jl}+2i\tilde{J}{\cal P}_{jl}}{\alpha_{jl}+2i\tilde{J}}.
\end{eqnarray*}
Choosing 
\begin{eqnarray}
\tilde{J} = \frac{J}{1+J^2},
\end{eqnarray}
allows us to express the $S$-matrix as,
\begin{eqnarray}
S_{jl} = \frac{\lambda_j-\lambda_l+ic
P_{jl}}{\lambda_j-\lambda_l+ic}  \frac{\lambda_j-\lambda_l-ic
{\cal P}_{jl}}{\lambda_j-\lambda_l-ic}.  \label{ees}
\end{eqnarray}
The S-matrices (\ref{eis},\ref{ees})
satisfy the Yang-Baxter conditions (\ref{ybe}), and also
\begin{eqnarray*}
S^{ij}S^{ik}S^{jk} = S^{jk} S^{ik} S^{ij},
\end{eqnarray*}
assuring that we were able to
 generate a cut-off version of the Hamiltonian while maintaining
integrability.

The cut-off scheme we introduced generates a flavor
component in the electron-electron $S$-matrix. Clearly
it captures the interaction among electrons induced by the impurity.
More so,  already  for the free
hamiltonian ${\cal H}_0 = -i \sum_{a}^{N} \sum_{m}^{f}
\int_{-\infty}^{\infty} dx 
\psi^{\dagger}_{a,m}(x)\partial_x \psi_{a,m}(x) $
 a non trivial S-matrix {\it must} be introduced if we choose an 
$SU(N) \times SU(f)$ invariant basis (which is appropriate for a subsequent 
inclusion of an impurity interaction)
rather than the simpler $SU(fN)$. A careful counting
 of states can be carried out \cite{destri}
to show that all expected states then appear with the correct degeneracies. 
It is instructive that this would  not be the case for the naive choice 
$S^{ij}_{(flavor)}=I$.

The energy eigenvalues of a $N^e$-electron state are a generalization of the
first line of (\ref{amp3}). They are of the form
\begin{eqnarray}
E = \sum_{j=1}^{N^e} k_j(1-\frac{k_j}{2\Lambda}).
\end{eqnarray}

\subsection[Eigenvalue equations]{Eigenvalue equations}

In order to determine the spectrum, we impose periodic
boundary conditions, and solve the corresponding eigenvalue problem. 
The procedure is standard \cite{lec} and we skip here the details.
The result is contained in
the Bethe Ansatz Equations (BAE) which we proceed to write down. 
Each of the degrees of freedom - charge, spin and flavor -
is described by a set of variables whose number depends on the symmetry
of the particular state.
The
charge degrees of freedom are given by the set
$\{k_j,j=1,...,N^e\}$. The spin degrees of
freedom are parameterized by the sets $\{\chi _\gamma ^r,\gamma
=1,...,M^r;r=1,...,N ;M^N =0\}$. Finally, the flavor degrees of freedom
are represented by the sets $\{\omega _\gamma ^r,\gamma =1,...,\bar{M}%
^r;r=1,...,f ;\bar{M}^f =0\}$. The set of integers $M^r; r=1,...,N-1$
specify the symmetry of the spin component of the wave function
given by a $SU(N)$  Young tableau with the length $l^r$ of the $r$th row 
given by $l^r=M^r-M^{r+1}, ~M^N =0, ~M^0=N^e+1 $. Similarly, the quantum numbers
$\{\bar{M}^r\}$ specify the symmetry of the  flavor component.

 The equations are 
\begin{eqnarray*}
\lefteqn{e^{ik_jL} =\prod_{\gamma =1}^{M^1}\frac{\chi _\gamma ^1-(1-\lambda
_j)+i\frac c2}{\chi _\gamma ^1-(1-\lambda _j)-i\frac c2}\prod_{\gamma =1}^{%
\bar{M}^1}\frac{\omega _\gamma ^1-\lambda _j+i\frac c2}{\omega _\gamma
^1-\lambda _j-i\frac c2}} & & \\
&-&\prod_{\beta =1}^{\bar{M}^r}\frac{\omega _\gamma ^r-\omega _\beta ^r+ic}{%
\omega _\gamma ^r-\omega _\beta ^r-ic} =\prod_{t=r\pm 1}\prod_{\beta =1}^{%
\bar{M}^t}\frac{\omega _\gamma ^r-\omega _\beta ^t+i\frac c2}{\omega _\gamma
^r-\omega _\beta ^t-i\frac c2}; \\ & & r=2,...,f -1, \\
&-&\prod_{\beta =1}^{\bar{M}^1}\frac{\omega _\gamma ^1-\omega _\beta ^1+ic}{%
\omega _\gamma ^1-\omega _\beta ^1-ic} = \prod_{j=1}^{N^e}\frac{\omega
_\gamma ^1-\lambda _j+i\frac c2}{\omega _\gamma ^1-\lambda _j-i\frac c2}%
\\ &\times & \prod_{\beta =1}^{\bar{M}^2}\frac{\omega _\gamma
^1-\omega _\beta ^2+i\frac 
c2}{\omega _\gamma ^1-\omega _\beta ^2-i\frac c2}, \\
&-&\prod_{\beta =1}^{M^r}\frac{\chi _\gamma ^r-\chi _\beta ^r+ic}{\chi _\gamma
^r-\chi _\beta ^r-ic} = \prod_{t=r\pm 1}\prod_{\beta =1}^{M^t}\frac{\chi
_\gamma ^r-\chi _\beta ^t+i\frac c2}{\chi _\gamma ^r-\chi _\beta ^t-i\frac c2%
}; \\ & & r=2,...,N-1, \\
&-&\prod_{\beta =1}^{M^1}\frac{\chi _\gamma ^1-\chi _\beta ^1+ic}{\chi
_\gamma 
^1-\chi _\beta ^1-ic} = \frac{\chi _\gamma ^1+i\frac c2}{\chi _\gamma
^1-i\frac c2}\prod_{j=1}^{N^e}\frac{\chi _\gamma ^1-(1-\lambda _j)+i\frac c2%
}{\chi _\gamma ^1-(1-\lambda _j)-i\frac c2} \\
&\times & \prod_{\beta =1}^{M^2}\frac{\chi
_\gamma ^1-\chi _\beta ^2+i\frac c2}{\chi _\gamma ^1-\chi _\beta ^2-i\frac c2%
}. 
\end{eqnarray*}

The next step is to solve the equations for all possible states,
identify the ground state and the low energy
 excitations. Subsequently, by summing over all excitation energies
we shall obtain the partition function.

The BAE contain the cutoff $\Lambda$ which
eventually is sent to infinity. We shall find that in this
limit  the equations reduce to a smaller set once the correct ground
state has been identified. It is composed of {\it string}-solutions (see below)
corresponding to electron composites
which interact most efficiently with the impurity. To sharpen our 
intuition we begin by some strong coupling considerations.

\subsubsection[Casimirology]{Casimirology}

As mentioned, the 
mechanism underlying the physics of the multichannel Kondo model
is the dynamic formation of electron composites. We expect that those
configurations are favored which allow minimization of the local
interaction at the impurity site. Consider then the general problem
of finding the ground state of the following hamiltonian:
\begin{eqnarray}
J \sum_{a}^{N^2-1} T^{(e)}_a T^{(i)}_a 
\label{casi1}
\end{eqnarray}
where the set $\{T^{e}_a, a=1,...,N^2-1\}$ is an arbitrary representation
of $SU(N)$, and $\{T^i_a, a=1,...,N^2-1\}$ is the particular
 representation of the impurity, in our case it will typically be
$(\mu)$.
In this article we will consider
impurities with spin in a totally symmetric representation (for more general
representations see \cite{zinn}). Each set is
normalized: $Tr(T_a T_b) = \frac{1}{2} \delta_{a,b}$.

The largest number of electrons allowed at the origin by the exclusion
principle is $N\times f$. This is obtained by placing $N$ electrons in
each of the channels. However, such a state is a singlet, both in spin
and in flavor, and gives a zero contribution to
(\ref{casi1}). Therefore, the number of electrons that form the composite,
$M$, is such that $M \le (N-1) \times f$. We will show here that if
the impurity is in a totally symmetric representation, the electron
composite that minimizes (\ref{casi1}) is made out of $M=(N-1)\times
f$ electrons. 

We will characterize the different representations of $SU(N)$ by their
Young tableaux. The fundamental representation is denoted by a box,
$\Box$, and the singlet by a point $\bullet$. The totally (anti)symmetric
representation resulting from the  
direct product of $\mu$ representations is denoted by a single
(column)row made out of $\mu$ boxes, where in the antisymmetric case we
assume $\mu \le N$.

\begin{picture}(200,100)(0,0)
\put(0,70){Antisymmetric}
\put(20,50){\vector(0,1){10}}
\put(19,40){$\mu$}
\put(20,30){\vector(0,-1){10}}
\put(45,20){\framebox(10,40){}}
\put(45,30){\line(1,0){10}}
\put(45,50){\line(1,0){10}}
\put(50,35){.}
\put(50,40){.}
\put(50,45){.}
\put(125,70){Symmetric}
\put(120,50){\framebox(60,10){}}
\put(130,50){\line(0,1){10}}
\put(170,50){\line(0,1){10}}
\put(140,55){.}
\put(150,55){.}
\put(160,55){.}
\put(147,40){$\mu$}
\put(140,45){\vector(-1,0){20}} 
\put(160,45){\vector(1,0){20}} 
\end{picture} 

An arbitrary representation resulting from a product of $M$ fundamentals
is associated with a Young tableau made up of  $M$ boxes, 
distributed in $k \le N$ rows. Let $m_j$ be the number of
boxes in the $j$-th row. Then, $m_j\ge m_{j+1}$, $\sum_{j=1}^{k} m_j =
M$. The corresponding Young tableau will be of the form

\begin{picture}(100,100)(0,0)
\put(50,80){\line(1,0){70}}
\put(50,70){\line(1,0){70}}
\put(50,60){\line(1,0){50}}
\put(50,50){\line(1,0){50}}
\put(50,40){\line(1,0){50}}
\put(50,30){\line(1,0){50}}
\put(120,70){\line(0,1){10}}
\put(130,70){$m_1$}
\put(110,70){\line(1,0){10}}
\put(110,60){\line(0,1){10}}
\put(100,60){\line(1,0){10}} 
\put(130,60){$m_2$}
\put(100,30){\line(0,1){30}}
\put(80,30){\line(1,0){20}}
\put(80,20){\line(0,1){10}}
\put(50,20){\line(1,0){30}}
\put(50,20){\line(0,1){60}}
\put(60,20){\line(0,1){60}}
\put(70,20){\line(0,1){60}}
\put(80,20){\line(0,1){60}}
\put(90,30){\line(0,1){50}}
\put(100,60){\line(0,1){20}}
\put(110,70){\line(0,1){10}}
\put(132,50){.}
\put(132,40){.}
\put(132,30){.}
\put(130,20){$m_k$}
\end{picture}

When $N=2$, the interaction (\ref{casi1}) can be written in terms of
conserved quantities 
\begin{eqnarray}
\lefteqn{J \vec{S}^{Me} \cdot \vec{S}^i =} \nonumber  & & \\ & &
\frac{J}{2} 
\left(S^{tot}(S^{tot}+1) 
- S^{Me}(S^{Me}+1)-S^{i}(S^{i}+1)\right). 
\label{casi2}
\end{eqnarray}
The operator $S(S+1)$ is a particular case of the Casimir operator,
$C(\Gamma)$, of $SU(N)$, which commutes with all the generators of the
group. For arbitrary $N$,
\begin{eqnarray}
J \sum_{a} T_a^{M} T_a^{i} = \frac{J}{2} \left( C(\Gamma^T) -
C(\Gamma^M) -  C(\Gamma^i) \right).
\end{eqnarray}
Given a representation $\Gamma$ of $SU(N)$ with $M$ boxes distributed
according to the set $\{m_j\}$, we have \cite{casi}
\begin{eqnarray}
C(\Gamma) = \frac{M(N^2-M)}{2N} + \frac{1}{2} \sum_{j=1}^{k} m_j
(m_j+1-2j). 
\end{eqnarray}
We can use Young tableaux as an easy way of decomposing the direct
product of representations into a direct sum. The procedure is
standard (see for instance \cite{Jones}). 

The electrons will be evenly distributed among the channels, forming a
flavor singlet. Hence, the spin of the electron composites is
described by a rectangular  tableau with $f$ columns and $k \le
N-1$ rows. Multiplying the electron tableau by the impurity tableau
we have (in the graphic representation we drop from the tableaux the singlet
 part consisting of columns of length $N$),

{
\setlength{\unitlength}{.2mm}
\begin{picture}(100,300)(0,0)
\put(10,130){\framebox(40,30){}}
\put(20,130){\line(0,1){30}}
\put(30,130){\line(0,1){30}}
\put(40,130){\line(0,1){30}}
\put(10,140){\line(1,0){40}}
\put(10,150){\line(1,0){40}}
\put(55,150){$\bigotimes$}
\put(80,150){\framebox(40,10){}}
\put(90,150){\line(0,1){10}}
\put(100,150){\line(0,1){10}}
\put(110,150){\line(0,1){10}}
\put(-5,145){$k$}
\put(27,168){$f$}
\put(97,168){$\mu$}
\put(135,150){\vector(1,0){30}}
\put(135,150){\vector(1,2){30}}
\put(135,150){\vector(1,-2){30}}
\put(210,250){\line(1,0){60}}
\put(210,240){\line(1,0){50}}
\put(210,230){\line(1,0){40}}
\put(210,220){\line(1,0){40}}
\put(270,240){\line(0,1){10}}
\put(210,210){\line(0,1){40}}
\put(220,210){\line(0,1){40}}
\put(230,210){\line(0,1){40}}
\put(240,210){\line(0,1){40}}
\put(250,210){\line(0,1){40}}
\put(260,240){\line(0,1){10}}
\put(250,210){\line(0,1){30}}
\put(210,210){\line(1,0){40}}
\put(250,240){\line(1,0){20}}
\put(237,262){$\mu$}
\put(227,195){$f$}
\put(162,225){$k+1$}
\put(280,225){$+ \cdots$}
\put(320,225){,  $\mu > f$}
\put(210,120){\framebox(40,40){}}
\put(220,120){\line(0,1){40}}
\put(230,120){\line(0,1){40}}
\put(240,120){\line(0,1){40}}
\put(210,130){\line(1,0){40}}
\put(210,140){\line(1,0){40}}
\put(210,150){\line(1,0){40}}
\put(162,135){$k+1$}
\put(227,167){$\mu$}
\put(280,135){$+ \cdots$}
\put(320,135){,  $\mu=f$}
\put(210,80){\line(1,0){40}}
\put(210,70){\line(1,0){40}}
\put(210,60){\line(1,0){40}}
\put(210,50){\line(1,0){30}}
\put(210,40){\line(0,1){40}}
\put(220,40){\line(0,1){40}}
\put(230,40){\line(0,1){40}}
\put(240,50){\line(0,1){30}}
\put(210,40){\line(1,0){30}}
\put(240,40){\line(0,1){10}}
\put(240,50){\line(1,0){10}}
\put(250,50){\line(0,1){30}}
\put(220,25){$\mu$}
\put(227,91){$f$}
\put(162,55){$k+1$}
\put(280,55){$+ \cdots$}
\put(320,55){,  $\mu < f$}
\end{picture}}

for $k < N-1$. If $k=N-1$, we have

{
\setlength{\unitlength}{.2mm}
\begin{picture}(350,220)(-50,0)
\put(10,100){\framebox(40,30){}}
\put(10,110){\line(1,0){40}}
\put(10,120){\line(1,0){40}}
\put(20,100){\line(0,1){30}}
\put(30,100){\line(0,1){30}}
\put(40,100){\line(0,1){30}}
\put(60,120){$\bigotimes$}
\put(90,120){\framebox(40,10){}}
\put(120,120){\line(0,1){10}}
\put(100,120){\line(0,1){10}}
\put(110,120){\line(0,1){10}}
\put(-50,115){$N-1$}
\put(27,138){$f$}
\put(97,138){$\mu$}
\put(140,120){\vector(1,0){30}}
\put(140,120){\vector(1,2){30}}
\put(140,120){\vector(1,-2){30}}
\put(220,170){\framebox(30,10){}}
\put(230,170){\line(0,1){10}}
\put(240,170){\line(0,1){10}}
\put(222,190){$\mu-f$}
\put(300,170){,  $\mu > f$}
\put(260,170){$+ \cdots$}
\put(235,116){$\bullet$}
\put(300,115){,  $\mu=f$}
\put(260,115){$+ \cdots$}
\put(220,30){\framebox(30,30){}}
\put(230,30){\line(0,1){30}}
\put(240,30){\line(0,1){30}}
\put(220,40){\line(1,0){30}}
\put(220,50){\line(1,0){30}}
\put(222,69){$f-\mu$}
\put(161,40){$N-1$}
\put(260,40){$+ \cdots$}
\put(300,40){,  $\mu < f$}
\end{picture}}

Notice that we have drawn only the terms in the
decompositions  that
give the lowest energy. The energy for all such configurations 
is given by 
\begin{eqnarray}
J \sum_a T_a^M T_a^i = - k \frac{min(\mu,f)}{2N}(N+max(\mu,f)).
\end{eqnarray}
Therefore, the energy is minimized when composites of $(N-1) \times f$
electrons are formed.

There are three different situations depending on the
value of $\mu/f$, as in the multichannel Kondo problem
\cite{Nozieres}. When $\mu > f$, there is underscreening: the
electrons cannot screen the impurity completely and the spin
configuration is characterized by a Young tableau with one row and
$\mu-f$ columns. As in the $N=2$ case, we will see later that such
object behaves as a free spin in the Kondo problem. The second case,
$\mu=f$, corresponds to complete screening: the 
electrons and the impurity form a singlet. This is a stable fixed
point of the full Hamiltonian with Fermi liquid behavior. Finally,
$\mu > f$ corresponds to overscreening: there are more electrons than
necessary  to screen the impurity. The resulting object 
corresponds to
 a tableau with $N-1$ columns and $f-\mu$ rows. This
configuration is unstable to the kinetic term, and the fixed point in
this case is characterized by Non-Fermi liquid behavior, as we will
see later. 

\subsubsection[Fusion]{Fusion}

We turn now to the dynamics of the full model captured by the BAE.
We shall argue that the ground state and low lying excitations lie
in a sector of the theory given by solutions of a particular form -
$f-strings$. Solutions of this type are $SU(f)$ flavor singlets -
 allowing them to have maximally large $SU(N)$ spin. We shall find 
that this class of excitations is characterized by a scale 
$T_0=De^{-\frac{2\pi }{N c}}$. When strings are broken to form
flavored excitations we expect them to be characterized by other
scales which will tend to infinity as the cut-off is removed and
thus not contribute to the impurity dynamics \cite{destri}.

The formation of composites in flavor corresponds to solutions of the
BAE where
the charge parameters, 
$\{\lambda_j\}$, are complex numbers centered around $\{\omega
_\gamma ^1\}$, according to the string hypothesis
\cite{afl,tw}. Likewise, rank $r$ 
flavor parameters are themselves centered around rank $r+1$ solutions
\cite{adkondo}. The
form of the charge parameters is, 
\begin{eqnarray*}
\lambda _\delta ^q=\frac{p_\delta }\Lambda +ic\left( \frac{f +1}2-q\right)
,~~~ q=1,2,...,f ,~~~ p_\delta ~~{real}. 
\end{eqnarray*}
while the flavor parameters,
\begin{eqnarray*}
\{\omega^r_{\gamma}, \gamma=1,2,..., M^r\} &=& \{p_A/\Lambda + 
iJ[(f-r+1)/2 -q]; \\ &&
 q = 1,2,...,f-r, A=1,...,N\}
\end{eqnarray*} 
where $r=0,1,...,f-1$. These configurations satisfy the BAE in a trivial manner
and induce {\it fusion} in the BAE equations as well as in the form of the
wavefunctions. A string built on momentum $p$ as its real part induces in the 
wave function a composite of the form $exp\{-\frac{1}{2}\Lambda J  
\sum_{j,l} |x_j-x_l| +
ip(x_1+...  +x_f)\} \times [ ...] $, which becomes local as $\Lambda 
\rightarrow \infty$. 

Inserting the string configurations into the full BAE we obtain the effective 
equations
governing the impurity spin dynamics. After removing the cutoff they become,
\begin{eqnarray*}
e^{if p_\delta L} &=&\prod_{\gamma =1}^{M^1}\frac{\chi _\gamma
^1-1+if\frac c 2}{\chi _\gamma ^1-1-if\frac c 2}, \\
-\prod_{\beta =1}^{M^r}\frac{\chi _\gamma ^r-\chi _\beta ^r+ic}{\chi _\gamma
^r-\chi _\beta ^r-ic} &=&\prod_{t=r\pm 1}\prod_{\beta =1}^{M^t}\frac{\chi
_\gamma ^r-\chi _\beta ^t+i\frac c2}{\chi _\gamma ^r-\chi _\beta ^t-i\frac c2%
}; \\ & & ~~~ r=2,...,N -1, \\ 
-\prod_{\beta =1}^{M^1}\frac{\chi _\gamma ^1-\chi _\beta ^1+ic}{\chi _\gamma
^1-\chi _\beta ^1-ic} &=&\frac{\chi _\gamma ^1+i\frac c2}{\chi _\gamma
^1-i\frac c2}\prod_{\delta =1}^{N^e/f }\frac{\chi _\gamma ^1-1+if\frac c
2}{\chi _\gamma ^1-1-if\frac c 2} \\ &\times & 
\prod_{\beta =1}^{M^2}\frac{\chi _\gamma
^1-\chi _\beta ^2+i\frac c2}{\chi _\gamma ^1-\chi _\beta ^2-i\frac c2},
\end{eqnarray*}
and the energy is given by 
\begin{eqnarray*}
E=\sum_{\delta =1}^{N^e/f }f p_\delta . 
\end{eqnarray*}

We proceed now to discuss the solutions of the fused equations. 
The solutions for the rank $r$ spin variables $\{\chi^r\}$ again fall into 
strings of arbitrary length $n$,
\begin{eqnarray*}
\chi _{\gamma ,j}^{r,n}=\chi _\gamma ^{r,n} \! \! +i\frac c2(n+1-2j);~
j=1,...,n,~ n=1,...,\infty . 
\end{eqnarray*}
and a state is characterized by the quantum numbers $M^{r,m}$ 
specifying the number of length-$m$ strings of rank $r$, ($%
\sum_{m=1}^\infty mM^{r,m}=M^r$).

 The equations coupling the real part of the strings, after summing
over the complex variables, can be conveniently written down in a logarithmic
 form. Let us first introduce the following
definitions: 
\begin{eqnarray*}
\theta _n(x) &\equiv &-2\arctan \left( \frac 2{nc}x\right) , \\
\phi _{n,m}^k(x) &\equiv &\sum_{j=1}^{\min (n,m)}\theta _{m+n+k-2j}(x), \\
\phi _{n,m}^0(x) &\equiv &\sum_{j=1}^{\min (n-1,m-1)}\theta _{m+n-2j}(x).
\end{eqnarray*}
Then, after some manipulations the Bethe-Ansatz equations take the form,
\begin{eqnarray*}
&&\sum_{m=1}^\infty \sum_{\beta =1}^{M^{r,m}}\left( \phi _{n,m}^2(\chi
_\gamma ^{r,n}-\chi _\beta ^{r,m})+\phi _{n,m}^0(\chi _\gamma ^{r,n}-\chi
_\beta ^{r,m})\right) \\
&=&2\pi I_\gamma ^{r,n}+\sum_{l=1}^\infty \sum_{\beta =1}^{M^{r-1,l}}\phi
_{n,l}^1(\chi _\gamma ^{r,n}-\chi _\beta ^{r-1,l}) \\ &+&
\sum_{l=1}^\infty 
\sum_{\beta =1}^{M^{r+1,l}}\phi _{n,l}^1(\chi _\gamma ^{r,n}-\chi _\beta
^{r+1,l}), \\
&&\sum_{m=1}^\infty \sum_{\beta =1}^{M^{1,m}}\left( \phi _{n,m}^2(\chi
_\gamma ^{1,n}-\chi _\beta ^{1,m})+\phi _{n,m}^0(\chi _\gamma ^{1,n}-\chi
_\beta ^{1,m})\right) \\
&=&2\pi I_\gamma ^{1,n}+\phi _{n,1}^1(\chi _\gamma ^{1,n})+\frac{N^e}f
\phi _{n,f }^1(\chi _\gamma ^{1,n}-1) \\ &+& \sum_{l=1}^\infty
\sum_{\beta 
=1}^{M^{2,l}}\phi _{n,l}^1(\chi _\gamma ^{1,n}-\chi _\beta ^{2,l}).
\end{eqnarray*}

The expression for the 
energy of the spin and charge sector is given by
\begin{eqnarray*}
E &=&\sum_{\delta =1}^{N^e/f }\frac{2\pi }{L} m_\delta  \nonumber \\
&+&\frac Df
\sum_{n=1}^\infty \sum_{\beta =1}^{M^{1,n}}\left( \phi _{n,f }^1(\chi
_\gamma ^{1,n}-1) - \pi \min (n,f )\right)
\end{eqnarray*}
where  $D=\frac{N^e}L$ is the electron density. It will turn out also to play the role of the cut-off.
In the presence of a magnetic field, $H$, there is a contribution to the
energy of the form,
\begin{eqnarray*}
 -2H\sum_{k=0}^{N-1}&& (M^k-M^{k+1}) ({\textstyle N-1\over 2}-k)= \\
&&= -2H \Bigl({\textstyle N-1\over 2}(N^e+1) -\sum_{r=1}^N M^r\Bigr).
\end{eqnarray*}

We now take the thermodynamic limit, $N^e\rightarrow \infty $, $L\rightarrow
\infty $, holding $D$ finite. In the limit we may replace sums with 
integrals after 
introducing densities of solutions, $\sigma _n^r(\chi )$, and densities of
holes in the distribution of solutions, $\sigma _n^{r,h}(\chi )$. The energy
is now written as 
\begin{eqnarray*}
E &=&E_c-H(N -1)(N^e+1) \\ &+& \sum_{n=1}^\infty\sum_{r=1}^{N-1} \int_{-\infty
}^\infty d\chi 
\sigma _n^r(\chi )g_{r,n}(\chi ), \\
\end{eqnarray*}
where  we introduced the energy function,
\begin{eqnarray*}
g_{r,n}(\chi ) &=&\frac Df \left( \phi _{n,f }^1(\chi -1)-\pi \min
(n,f )\right) \delta _{r,1}+2Hn, 
\end{eqnarray*}
and $E_c$  denotes the contribution of the charge sector to the energy,
\begin{eqnarray*}
E_c &=&\sum_{\delta =1}^{N^e/f }\frac{2\pi }Lm_\delta.
\end{eqnarray*}

In the thermodynamic limit the BAE are replaced by  integral equations for 
 the densities, $\{\sigma _n^r,\sigma _n^{r,h}\}$. Standard manipulations
\cite{lec} lead to,
\begin{eqnarray*}
\sigma _n^{r,h}(\chi ) = -\sum_{m=1}^\infty
\sum_{s=0}^N \int_{-\infty }^\infty d\chi
^{\prime}A_{n,m}^{r,s}(\chi -\chi ^{\prime })\sigma _m^s(\chi ^{\prime
}).  
\end{eqnarray*}
where we introduced the following operators:
\begin{eqnarray*}
K_{n,m}^\alpha (\chi ) &\equiv &\sum_{j=1}^{\min (n,m)}k_{m+n+\alpha
-2j}(\chi ), \\
\tilde{K}_{n,m}^0(\chi ) &\equiv &\sum_{j=1}^{\min
(n-1,m-1)}k_{m+n-2j}(\chi ), \\ 
k_\alpha (\chi ) &\equiv &-\frac 1{2\pi }\theta _\alpha (\chi ),~~~
k_0(\chi )\equiv \delta (\chi ).
\end{eqnarray*}

and

\begin{eqnarray*}
& & A_{n,m}(\chi ) \equiv K_{n,m}^2(\chi )+K_{n,m}^0(\chi ),  \\
& & B_{n,m}(\chi ) \equiv K_{n,m}^1(\chi ) \\
& & A_{n,m}^{r,s} \equiv  A_{n,m}\delta ^{r,s}-B_{n,m}(\delta
^{r,s+1}+\delta^{r,s-1}),~~ r\ge 1, 
\end{eqnarray*}
and by convention,
\begin{eqnarray}
&&\sigma _l^0(\chi )\equiv \delta (\chi )\delta _{l,1}+\frac{N^e}f \delta
(\chi +1)\delta _{l,f }, \\
&& \sigma_n^N(\chi) \equiv 0\;.
\end{eqnarray}

We shall not analyze here the ground state and the individual excitations.
Instead, we shall proceed to derive the thermodynamic properties of the model.
 
\section[Thermodynamics]{Thermodynamics}

\subsection[Thermodynamic Bethe Ansatz equations]{Thermodynamic Bethe
Ansatz equations} 

We now calculate the impurity contribution to the free energy, using the
well known formalism of Yang and Yang\cite{afl,tw}. We seek to find the
 configuration $ \{\sigma_n^r(\chi )+\sigma
_n^{r,h}(\chi )\}$ which would extremize the free energy.
The entropy of such a configuration is,
\begin{eqnarray*}
{\cal S}=\sum_{n,r}\int d\chi &&\left\{ \left( \sigma_n^r(\chi )+\sigma
_n^{r,h}(\chi )\right) \ln \left( \sigma _n^r(\chi )+\sigma _n^{r,h}(\chi
)\right) \right. \\
&&\left. -\sigma _n^r(\chi )\ln \sigma _n^r(\chi )-\sigma _n^{r,h}(\chi )\ln
\sigma _n^{r,h}(\chi )\right\},
\end{eqnarray*}
and its contribution to the spin free energy
\begin{eqnarray*}
F &=&E-T{\cal S}= \\ && -H(N -1)(N^e+1)+\sum_{n,r}\int d\chi \left\{
\sigma 
_n^r(\chi )g_{r,n}(\chi ) \right. \\ &&-T\left(
\left( \sigma _n^r(\chi )+\sigma _n^{r,h}(\chi )\right) \ln \left( \sigma
_n^r(\chi )+\sigma _n^{r,h}(\chi )\right)  \right. \\
&&\left. \left. -\sigma _n^r(\chi )\ln \sigma _n^r(\chi )-\sigma
_n^{r,h}(\chi )\ln \sigma _n^{r,h}(\chi )\right) \right\}.
\end{eqnarray*}
The free energy is varied with respect to the densities, subject
to constraints imposed by the Bethe-Ansatz equations, 
\begin{eqnarray*}
\delta \sigma _n^{r,h}(\chi ) &=& -\sum_{m,s}\int_{-\infty }^\infty d\chi
^{\prime }A_{n,m}^{r,s}(\chi -\chi ^{\prime })\delta \sigma _m^s(\chi
^{\prime }), \\  \delta \sigma _n^0(\chi )&=& \delta \sigma _n^N (\chi
)=0.  
\end{eqnarray*}
We obtain the following infinite set of integral equations for
 the equilibrium densities,
\begin{eqnarray} \label{eqq}
\lefteqn{\ln (1+\eta _n^r(\chi ))=} && \label{tick}\\ && \frac{g_{r,n}(\chi
)}T+\sum_{m,s}\int_{-\infty 
}^\infty d\chi ^{\prime }A_{n,m}^{r,s}(\chi -\chi ^{\prime })\ln (1+(\eta
_m^s(\chi ^{\prime }))^{-1}), \nonumber
\end{eqnarray}
where 
\[
\eta _n^r(\chi )\equiv \frac{\sigma _n^{r,h}(\chi )}{\sigma _n^r(\chi )}%
,~~~ (\eta _n^N )^{-1}\equiv (\eta _n^0)^{-1}\equiv 0. 
\]
We transform this set of equations with the help of the 
following identities (we will now drop the functional dependence) 
\begin{eqnarray*}
\lefteqn{A_{n,m}^{r,s}-G(A_{n-1,m}^{r,s}+A_{n+1,m}^{r,s})} \\ &=& \delta
_{n,m}\delta 
^{r,s}-G\delta _{n,m}(\delta ^{r,s+1}+\delta ^{r,s-1}) \\
& & A_{1,m}^{r,s}-GA_{2,m}^{r,s} = \delta _{1,m}\delta ^{r,s}-G\delta
_{1,m}(\delta ^{r,s+1}+\delta ^{r,s-1}),
\end{eqnarray*}
with the integral operator $G$ defined as 
\begin{eqnarray*}
Gf(\chi ) \equiv \frac{[1]}{[0]+[2]}f(\chi )\equiv \frac
1{2c}\int_{-\infty }^\infty d\chi ^{\prime }\frac{f(\chi ^{\prime })}{\cosh
\left( \frac \pi c(\chi -\chi ^{\prime })\right) }, 
\end{eqnarray*}
and 
$\lbrack n]f(\chi ) \equiv \int_{-\infty }^\infty d\chi ^{\prime }k_n(\chi
-\chi ^{\prime })f(\chi ^{\prime })$.
We find,
\begin{eqnarray}
\ln \eta _1^r &=&-\frac{2}{f}\frac D{T}\arctan e^{\frac \pi c(\chi
-1)}\delta 
^{r,1}\delta _{n,f }+G\ln (1+\eta _1^r)  \nonumber \\
&&-G(\ln (1+(\eta _1^{r-1})^{-1})+\ln (1+(\eta _1^{r+1})^{-1})),
\label{easy} \\
\ln \eta _n^r &=&-2\frac D{Tf }\arctan e^{\frac \pi c(\chi -1)}\delta
^{r,1}\delta _{n,f }  \nonumber \\
&&+G(\ln (1+\eta _{n-1}^r)+\ln (1+\eta _{n+1}^r))  \nonumber \\
&&-G(\ln (1+(\eta _n^{r-1})^{-1})+\ln (1+(\eta _n^{r+1})^{-1})),
\label{easy2}
\end{eqnarray}
with boundary conditions,  

\begin{eqnarray}
 \lim_{n\rightarrow \infty }\left( [n+1]\ln (1+\eta
_n^r)-[n]\ln (1+\eta_{n+1}^r)\right) 
= -2\frac HT.  \label{bc}
\end{eqnarray}
which follow directly from eqns (\ref{eqq}).
Another form of these equation can be obtained after inverting, see
 Ref.~[\onlinecite{afl}],
\begin{eqnarray}
&-& \ln (1+(\eta _n^r)^{-1}) = -\frac{\Delta E_{N
,r}^{fund}}T\delta _{n,f} \nonumber  \\ 
&+& \sum_{q=1}^{N -1}G_N ^{r,q}\left( \ln
(1+\eta _{n+1}^q) + \ln (1+\eta_{n-1}^q)  \right. \nonumber  \\ 
&-& \left. G^{-1}\ln (1+\eta _n^q)\right) , 
\label{label}
\end{eqnarray}
where $\ln (1+\eta _0^r)\equiv 0$, and 
 the Fourier transform of the kernel of the integral operator $G_N
^{r,q}$ is given by
\[
\widetilde{G}_N ^{r,q}(p)\equiv \frac{\sinh \left( \min (r,q)\frac{cp}%
2\right) \sinh \left( (N -\max (r,q))\frac{cp}2\right) }{\sinh \left( 
\frac{cp}2\right) \sinh \left( N \frac{cp}2\right) }. 
\]
The driving term in these equations,

\[
\Delta E_{N ,r}^{fund}=G_N ^{r,1}G^{-1}\left( 2\frac Df \arctan
e^{\frac \pi c(\chi -1)}\right)
\]
is the energy of the fundamental excitation. It can be calculated explicitly,
\begin{eqnarray*}
& & \Delta E_{N ,r}^{fund}=\frac Df \left\{ \pi \frac{N -r}N \right. \\
&-& \!
\left. 2\arctan \! 
\left(\! \tan \left( \frac \pi 2\frac{N -r}N \right) \tanh \! \left(
\frac \pi 
{N c}(\chi -1)\right)\! \right) \! \right\} . 
\end{eqnarray*}
\begin{figure}
\epsfxsize=8cm
\hskip0.5cm
\epsfbox{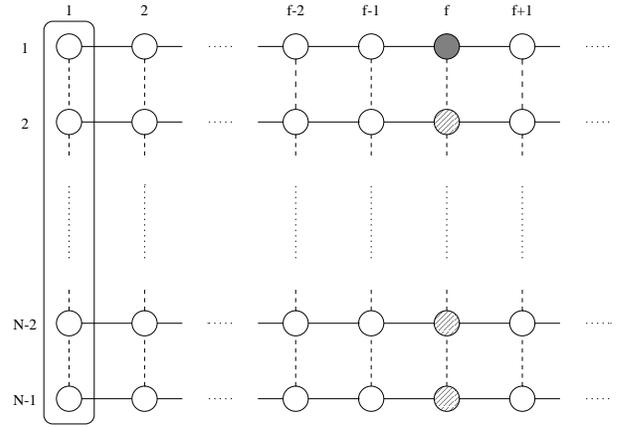}
\vskip0.7truecm  
\caption{Diagrammatic representation of the integral equations. The
circles correspond to the functions $\eta_n^r$. The filled circle
indicates that the equation for the corresponding $\eta_n^r$ has a
driving term. The circles with stripes indicate that the corresponding
$\eta_n^r$ have  driving terms in the other set of TBA equations. The
solid line indicates a link between two $\eta_n^r$ 
thorough the convolution $G\ln(1+\eta)$. The dashed line indicates a
link through $G\ln (1+1/\eta) $. Finally, the box encircles the
$\eta_1^r$, which are the functions used to evaluate the impurity
contribution to the free energy.} 
\label{fig1}
\end{figure}
A pictorial description of (\ref{label}) is shown in
Fig. \ref{fig1}. The circles correspond to the functions $\eta_n^r(x)$,
 and are arranged according to their indices. The lines
join functions that appear in the same equation. The  full and dotted
lines  indicate that the functions $\eta_n^{r\pm 1}$ and $\eta_{n\pm 1}^r$
appear  differently in 
eqns (\ref{easy},\ref{easy2}). The driving terms in (\ref{easy},\ref{easy2})
correspond to filled dark circles in Fig. \ref{fig1}. The diagonally lined
circles correspond to functions 
associated to a driving term in (\ref{label}). Clearly
there are two regions: the one, $n<f$, contains a finite number
of functions $\eta_n^r$, while the other, $n>f$, is unbounded. The regions
 are  separated
by the column with the driving terms, $n=f$. When studying the
low-T properties of the system we will only need to consider one 
region at a time.

We will now write the free energy in terms of the set $\{\eta_n^r\}$. Using
the integral equations for the densities we can write 
\begin{eqnarray*}
F &=&F_0+\sum_{n,r}\int d\chi \left( g_{r,n}\sigma _n^r-T\sigma _n^r\ln
(1+\eta _n^r) \right. \\ &+& \left. T\sum_{m,s}\int d\chi ^{\prime
}A_{n,m}^{r,s}\sigma _n^r\ln 
(1+(\eta _m^s)^{-1})\right) 
\end{eqnarray*}
where $F_0 =E_c-H(N -1)(N^e+1)$ is the ground state energy.
After a few further manipulations the free energy can be written as, 
\begin{eqnarray*}
F &=& F_0-T\sum_n\int_{-\infty }^\infty d\chi \ln (1+(\eta
_n^1)^{-1}) \times \\
 && \times \Bigl\{ k_n(\chi ) +  \frac{N^e}f
\sum_{j=1}^{\min (n,f )}k_{f +n+1-2j}(\chi -1) \Bigr\}.
\end{eqnarray*}
We are only interested in the impurity contribution to the free
energy, $F^i $, which contains all the effects of the interaction. 
It is 
\begin{eqnarray*}
F^i &=& -T\sum_{n=1}^\infty
\int_{-\infty }^\infty d\chi \delta (\chi )[n]\ln (1+(\eta
_n^1)^{-1}) \\
&=& \sum_{q=1}^{N -1}\int_{-\infty }^\infty d\chi G_N ^{1,q}(\chi
)g_1^q(\chi ) \\ &-&T\sum_{q=1}^{N -1}\int_{-\infty }^\infty d\chi G_N
^{1,q}(\chi )\ln (1+\eta _1^q), 
\end{eqnarray*}
The first term corresponds to the impurity contribution to the ground
state. At finite temperatures we are only interested in the second term
which after further manipulations becomes,

\[
F^i=- T \sum_{q=1}^{N-1} \! \int_{-\infty }^\infty  \!\!\!\!\!\! d\chi
\frac 1{N 
c}
\frac{\sin \pi \frac{N -q}N }{\cosh \frac{2\pi \chi }{N c}+\cos \pi 
\frac{N -q}N }\ln (1+\eta _1^q(\chi )). 
\]
When the impurity is in the fundamental representation, only the
$\eta_1^r$ functions contribute to $F^i$. In Fig. \ref{fig1} this
feature corresponds to a box drawn around the first column.

\subsection[Scaling limit]{Scaling limit}

We will now take the scaling limit, $D\rightarrow \infty $,
$c\rightarrow 0$%
, $T_0$ constant, where 
\[
T_0=De^{-\frac{2\pi }{N c}}. 
\]
This is the correct limit as discussed in Ref.~[\onlinecite{afl}].
We also introduce the new variable 
\[
\xi =\frac{2\pi }{N c}\chi +\ln \frac{T_0}T, 
\]
so 
\begin{eqnarray}
F^i=-\frac T{2\pi }\sum_{q=1}^{N-1}\int_{-\infty }^\infty d\xi 
\frac{\left(\sin \pi \frac{N -q}N\right) \ln (1+\eta _1^q(\xi ))
}{\cosh \left( \xi -\ln \frac{T_0}T\right) 
+\cos \pi \frac{N -q}N }.
\label{free}
\end{eqnarray}
The only modification in the thermodynamic equations is in $\Delta E_{N
,r}^{fund}$. Thus
\begin{eqnarray}
&-&\ln (1+(\eta _n^r)^{-1}) =-\frac{2}{f} e^\xi \sin \left( \frac{\pi
r}N \right) 
\delta _{n,f } \label{tba} \\ &+& \sum_{q=1}^{N -1}G_N
^{r,q}( \ln (1+\eta 
_{n+1}^q)+\ln (1+\eta _{n-1}^q) \nonumber \\ &-& G^{-1}\ln (1+\eta
_n^q) ) 
, \nonumber \\ 
\eta_0^r
&\equiv& 0,~\ln(1+(\eta_n^0)^{-1}) \equiv
0,~\ln(1+(\eta_n^{f})^{-1}) \equiv 0,  
\end{eqnarray}
with boundary conditions 
\begin{eqnarray}
\lim_{n\rightarrow \infty } \!\! \left( [n+1]\ln (1+\eta _n^r)-[n]\ln
(1+\eta 
_{n+1}^r)\right) \!  = \! -2\frac HT \label{tba2}.
\end{eqnarray}

\subsection[Asymptotic solutions - low temperature properties]{Asymptotic
solutions-Low temperature properties} 

Here we will study several asymptotic limits of the thermodynamic
integral equations. Some technical points will be considered
in detail.
It is not easy to study analytically the integral equations
(\ref{tba}-\ref{tba2}) due to the complexity of the operator
$G_N^{r,q}$. 
Instead, we will study the equivalent set 
equations (\ref{easy}-\ref{bc}). 
We will discuss the appropriate procedure
to obtain the asymptotic solutions of the equations order by order.

 The zeroth order approximation yields a description of the fixed point
 itself, the corrections (first order) describe its neighborhood.

\subsubsection[Zeroth order- the fixed point ]{Zeroth order - the fixed point}

The functions $\eta_n^r$ tend either to 0 or to constant values as
the magnitudes of their arguments tend to infinity. The only
information  needed about the driving term is that it
tends to 0 as $\xi \rightarrow -\infty$, and  to $-\infty$ as
$\xi \rightarrow \infty$. Therefore, the parameter $\chi$ does not
appear explicitly in (\ref{easy}-\ref{bc}), and the kernel of
$G$  can be replaced by
$(1/2)\delta(\xi-\xi')$.

Thus, the zeroth order problem consists in evaluating the
following set of constants
\begin{eqnarray*}
\eta_n^{r,\pm} &\equiv & \eta_n^r(\xi \rightarrow \pm \infty).
\end{eqnarray*}
When $\xi \rightarrow -\infty$, all the driving terms vanish, and the
algebraic equations for the set $\{ \eta_n^{r,-} \}$ are
\begin{eqnarray*}
&&2 \ln \eta_n^{r,-} = \ln(1+\eta_{n+1}^{r,-}) + 
\ln(1+\eta_{n-1}^{r,-})  \\
&& \phantom{nnnnnnn}+ \ln(1+(\eta_n^{r+1,-})^{-1}) + 
\ln(1+(\eta_n^{r-1,-})^{-1}),
\\ 
&& \eta_0^{r,-} = 0, \\
&& \ln(1+(\eta_{n}^{0,-})^{-1}) = \ln(1+(\eta_n^{N,-})^{-1}) = 0, \\    
&&-2\frac HT =\lim_{n\rightarrow \infty }\left( [n+1]\ln (1+\eta
_n^{r,-})-[n]\ln (1+\eta_{n+1}^{r,-})\right), 
\end{eqnarray*}
Since the kernel of $G_N ^{r,q}(\chi )$ satisfies 
\begin{eqnarray*}
G_N ^{r,q} =G_N ^{q,r},~~ G_N ^{r,q} = G_N ^{N -r,N -q},
\end{eqnarray*}
we must have $\eta_n^{N-r} = \eta_n^r$. The solution 
is easily obtained as it is independent of the flavor symmetry, 
\begin{eqnarray}
\eta_{n}^{r,-} &=& \frac{\sinh((n+r)x_0) \sinh((n+N-r)x_0)}
{\sinh(rx_0) \sinh((N-r)x_0)}-1, \nonumber \\ & &
n=1,2,...,~~r=1,...,N-1,~~x_0 = \frac{H}{T}
\label{eta1}
\end{eqnarray} 
As for $\eta_n^{n,+}$, we should consider separately the cases $n>f$
and $n<f$ ($\eta_f^{r,+}=0$). In the first case, we have equations
similar to those for $\{\eta_n^{r,-}\}$, except that all the indices
$n$ are {\it shifted} by $f$, in analogy with the multichannel Kondo
model. Hence
\begin{eqnarray}
\eta_{n}^{r,+} &=& \frac{\sinh((n-f+r)x_0) \sinh((n-f+N-r)x_0)}
{\sinh(rx_0) \sinh((N-r)x_0)}-1, \nonumber \\ &&
n=f,f+1,...,~~~r=1,...,N-1.  \label{eta2}
\end{eqnarray} 
Finally, for $n<f$ there are a finite number of $\eta_n^{r,+}$
involved, since $\eta_f^{r,+}=0$, and $G\ln(1+\eta_f^r)=0$. As in the 
multichannel Kondo \cite{adkondo}, the $sinh$ functions are replaced by $sin$ 
functions, and the
coefficients are independent of the magnetic field. Thus
\begin{eqnarray}
\eta_n^{r,+} &=& \frac{\sin(\frac{\pi}{(f+N)} (n+r))
\sin(\frac{\pi}{(f+N)}(n+N-r))} 
{\sin(\frac{\pi}{(f+N)}r)
\sin(\frac{\pi}{(f+N)}(N-r))}-1, \nonumber \\
& & n=1,...,f-1~~~r=1,...,N-1. 
\label{eta3}
\end{eqnarray} 
These results coincide with those obtained 
in Ref.~[\onlinecite{tsvelik}] for a
model of interacting fermions with the same symmetry. This is not surprising
since these results depend only on the symmetry of the problem.  
Notice that the multichannel Kondo results correspond to
(\ref{eta1}-\ref{eta3}) with $N=2$. 
Some features of (\ref{eta1}-\ref{eta3}) can be appreciated in
Fig. \ref{fig1}. When $\xi \rightarrow -\infty$ the driving term does
not contribute to the equations and the situation is the same as in
the Coqblin-Schrieffer model. For $\xi \rightarrow \infty$, and $n \ge
f$, we can disregard the $n<f$ sector, and the leftover diagram is
effectively the same as for the Coqblin-Schrieffer model with the 
substitution $n \rightarrow n-f$. Finally, for $n<f$ we have a finite
number of $\eta_n^{r,+}$ involved. Hence the replacement of the
$\sinh$ by $\sin$. 

\subsubsection[Residual entropy - the fixed point ]{Residual entropy - the fixed point}

Here we will calculate the residual entropy in the overscreened case,
$f>1$. As $T \rightarrow 
0$, the dominant term in the free energy will be linear, and it will
depend only on the values of $\eta_1^{r,+}$.
\begin{eqnarray}
F^i &\sim & -\frac{T}{2\pi} \sum_{q=1}^{N-1} \sin\left(\pi
\frac{N-q}{N} \right) \\ &\times & \int_{0}^{\infty}d\xi
\frac{\ln(1+\eta_1^{q,+})}{\cosh  
(\xi-\ln \frac{T_0}{T}) + \cos\left(\pi \frac{N-q}{N} \right)} 
\label{F_imp}\\
&\sim & -T \sum_{q=1}^{N-1} \left( \frac{N-q}{N} \right) 
\ln(1+\eta_1^{q,+}),
\end{eqnarray}
Substituting the values of $\eta_1^{q,+}$, and
 taking advantage of the symmetry that $\eta_1^{N-q,+} =
\eta_1^{q,+}$, we find
\begin{eqnarray*}
F^i = -T \ln \frac{\sin \frac{\pi N}{f+N}}{\sin \frac{\pi}{f+N}}
\end{eqnarray*}
Hence, the residual entropy is 
\begin{eqnarray}
S^i_{T=0} = \left. -\frac{\partial F^i}{\partial T} \right|_{T=0} =\ln 
\frac{\sin \frac{\pi N}{f+N} }{\sin \frac{\pi}{f+N}}=\ln 
\frac{\sin \frac{\pi f}{f+N} }{\sin \frac{\pi}{f+N}}
\label{rsi}
\end{eqnarray}
Once again, we recover the multichannel results if we set $N=2$. It is
quite clear that it is not the logarithm of an integer number! 

The expression for the entropy can be written as the sum of two terms:
one that depends only on $N+f$ and a second one that depends only on
$|\log\gamma|$, ($\gamma=f/N$)
\begin{eqnarray}
S^i = \ln \sin \frac{\pi}{1+e^{|\log \gamma|}} - \ln \sin \frac{\pi}{N+f}
\label{rsi1}
\end{eqnarray}
In the limits $f \gg N$ and $f \ll N$ we have
\begin{eqnarray}
S^i = \left\{ \begin{array}{cl} \ln N -\frac{\pi^2}{6}
\frac{N^2-1}{f^2}, & f \gg N, \\ \\
\ln f -\frac{\pi^2}{6}
\frac{f^2-1}{N^2}, & f \ll N. \end{array} \right.
\label{rsi2}
\end{eqnarray}
Furthermore, it is clear from (\ref{rsi1}) that two systems
characterized by 
$\gamma_1$ and $\gamma_2$ such that $\gamma_1 = 1/\gamma_2$ have the
same residual entropy. When $\gamma=1$, the first term is zero.

\begin{figure}
\epsfxsize=8cm
\hskip0.5cm
\epsfbox{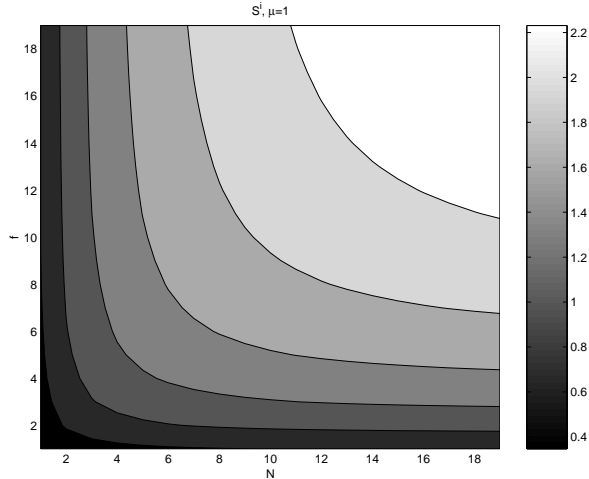}
\vskip0.7truecm  
\caption{Overscreening residual entropy, $S^i$, for an impurity in
the fundamental 
representation of $SU(N)$, and for different values of $N$ and $f$.}
\label{ent1}
\end{figure}

Fig. \ref{ent1} corresponds to (\ref{rsi}) for different values of $N$
and $f$. It is quite apparent that the value of $S^i$ increases with
$N+f$. It is also clear that the figure is symmetric with respect to
the $N=f$ axis, which means that $S^i$ is the same for $\gamma$ and
for $1/\gamma$. Finally, if we fix $N+f$, the largest value of the
residual entropy corresponds to $N=f$.  
In terms of the diagrammatic construction, Fig. \ref{fig1}, $S^i$
measures the size of the overscreened region and how asymmetric the
region is. For fixed $N+f$, the largest residual
entropy corresponds to $\gamma=1$, in the same way as the square is
the rectangle with the largest area for a fixed perimeter. This can be
seen in the following diagrams, were we have omitted the lines and
drawn only the circles corresponding to $n<f$.

\begin{picture}(100,140)(10,20)
\put(10,135){$N=7, f=3$}
\multiput(10,10)(0,20){6}{\multiput(10,10)(20,0){2}{\circle{10}}}
\end{picture}

\begin{eqnarray}
\label{p1}
\end{eqnarray}

\begin{picture}(100,120)(10,20)
\put(10,100){$N=f=5$}
\multiput(10,10)(0,20){4}{\multiput(10,10)(20,0){4}{\circle{10}}}
\end{picture}

\begin{eqnarray}
\label{p2}
\end{eqnarray}

\begin{picture}(100,70)(10,20)
\put(10,55){$N=3, f=7$}
\multiput(10,10)(0,20){2}{\multiput(10,10)(20,0){6}{\circle{10}}}
\end{picture}

\begin{eqnarray}
\label{p3}
\end{eqnarray}

The largest entropy corresponds to the configuration with the largest
number of circles, for a fixed value of $N+f$. That is, $N=f$. Notice
that the first and the third cases have the same number of circles
and, indeed, the same value of of the impurity entropy $S^i$.

\subsubsection[First order - the neighborhood of the fixed 
point ]{First order - the neighborhood of the fixed point}

Now we turn to the calculation of the thermodynamic properties
of the $SU(N)\times SU(f)$ model well below the Kondo scale $T_0$. 
As is obvious from Eq.~(\ref{F_imp}) for $T\ll T_0$ the nontrivial
temperature dependence of the impurity free energy is determined by the
asymptotic behavior of the functions $\eta^q_1(\xi)$ in the region $\xi\gg 1$.
Therefore
we want to find the dominant dependence of $\eta_n^r$ on $\xi$ for
$n<f$ and $\xi$ large and positive. To this purpose we will only need 
the equations with $n<f$, which do not have a driving term, and 
the asymptotic value of $\eta_f^r$.

Consider the action of the operator $G$
\begin{eqnarray*}
G_\chi f &=& \frac{1}{2c} \int_{-\infty}^{\infty} d\chi'
\frac{f(\chi')}{\cosh(\frac{\pi}{c} (\chi-\chi'))}  \\ &=&
\frac{1}{2c} \int_{-\infty}^{\infty} d\chi'
\frac{\tilde{f}(\frac{2\pi \chi'}{N c}+\ln \frac{T_0}{T})}
{\cosh(\frac{N}{2} \frac{2\pi}{N c} (\chi-\chi'))} \\  
&=& \frac{N}{4\pi} \int_{-\infty}^{\infty} d\xi'
\frac{\tilde{f}(\xi')}{\cosh(\frac{N}{2} (\xi-\xi'))} \equiv G_\xi
f,
\end{eqnarray*}
where we have dropped the tilde in the last expression. This
establishes the correspondence between functions and variables of the
two systems of equations. 

{}From the set of equations (\ref{easy}-\ref{bc}), and the asymptotic
values (\ref{tba}-\ref{tba2}) we learn that for large
and positive values of $\xi$ we have (if $x_0 = H/T$ is very small)
\begin{eqnarray}
\eta_f^r(\xi) \propto (\alpha^r+\beta^r x_0^2) e^{-\frac{2}{f}\sin \left(
\frac{\pi r}{N} \right) e^\xi}
\end{eqnarray}
We now evaluate the dominant contribution to $G_{\xi} \ln(1+\eta_f^r)$
\begin{eqnarray*}
&&G_{\xi} \ln(1+\eta_f^r) = {\textstyle \frac{N}{4\pi}} \int_{-\infty}^{\infty}
d\xi' \frac{\ln(1+\eta_f^r)}{\cosh \frac{N}{2} (\xi-\xi')} \\ \\ 
&& \phantom{nnnn} \stackrel{\xi\gg1}{\sim}{\textstyle \frac{N}{4\pi}} 
(\alpha^r+\beta^r
x_0^2)\int_{\omega}^{\infty} d\xi' 
\frac{e^{-\Delta e^{\xi'}}}{\cosh \frac{N}{2} \xi'},
\end{eqnarray*}
where 
$\omega$ is some lower cutoff of the integral of the order of unity and
\begin{eqnarray*}
\Delta \equiv \frac{2}{f} \sin \left(\frac{\pi r}{N} \right) e^\xi.
\end{eqnarray*}
For large $\xi$, the only relevant contribution to the integral occurs
around $e^{\xi'} \sim 1/\Delta$. Therefore, we approximate the
previous integral by
\begin{eqnarray*}
 (\alpha^r+\beta^r
x_0^2)\frac{\Delta^{\frac{N}{2}}}{1+\Delta^{N}} \sim 
(\alpha^r+\beta^r x_0^2) e^{-\frac{N}{2} \xi}.
\end{eqnarray*} 
Notice that this is correct up to terms of the form $\xi^{\mu}
e^{-\frac{N}{2} \xi}$ which cannot be accounted for using this crude
approximation.

The previous calculation indicates that $\eta_{n<f}^r$ will have a
contribution of order $e^{-\frac{N}{2} \xi}$, since $G_\xi
e^{-\alpha \xi} \propto e^{-\alpha \xi}$, as we will see
below. Therefore, we have to determine whether there are contributions
more singular still. In other words, we have to find out if there
are solutions of the integral equations for large $\xi$ of the form 
\begin{eqnarray*}
\eta_{n<f}^r(\xi) \sim \eta_{n<f}^{r,+} + c_n^r(\alpha^r+\beta^r
x_0^2) e^{-\tau \xi}, ~~with~~\tau < \frac{N}{2}.
\end{eqnarray*}

Introducing the eigenvalue $\lambda \equiv 2\cos \frac{\pi
\tau}{N}$, so that 
\begin{eqnarray*}
G e^{-\tau \xi} = \frac{e^{-\tau \xi}}{\lambda}
\end{eqnarray*}
we proceed to convert the TBA equations into algebraic recursion
relations. Noting that,

\begin{eqnarray*}
\ln(1+\eta_n^r) &\sim & \ln(1+\eta_n^{r,+})+(b_n^r+a_n^r x_0^2)  e^{-\tau
\xi}, \\
\ln \eta_n^r &\sim& \ln_n^{r,+} + \frac{(b_n^r+a_n^r x^2_0)}{\omega_n^r}
e^{-\tau \xi}
\end{eqnarray*}

where,

\begin{eqnarray*}
(b_n^r + a_n^r x_0^2)
&\equiv& \frac{c_n^r(\alpha^r+\beta^r x_0^2)}{1+\eta_n^{r,+}}\\
 \omega_n^r& \equiv& \frac{\eta_n^{r,+}}{1+\eta_n^{r,+}} 
\end{eqnarray*}

then substituting in the integral equations (\ref{easy},\ref{easy2}) for $n<f$ and 
using the zeroth-order results
we obtain the following set of algebraic equations for the coefficients
of $e^{-\tau \xi}$ (we only write the equations for $b_n^r$ since they
are identical to those for  $a_n^r$),
\begin{eqnarray*}
\lambda b_n^r = \omega_n^r (b_{n+1}^r+b_{n-1}^r) +
\frac{\omega_n^r}{\eta_n^{r+1,+}}b_n^{r+1} +
\frac{\omega_n^r}{\eta_n^{r-1,+}}b_n^{r-1}, 
\end{eqnarray*}
with
\begin{eqnarray}
b_0^r = b_f^r =0. \label{4}
\end{eqnarray}
More explicitly, upon     inserting zero order values the
equations become,
\begin{eqnarray}
\lambda b_n^r &=&
\frac{\sin((n+N)a)\sin(na)}{\sin((n+r)a)\sin((n+N-r)a)}
(b_{n+1}^r+b_{n-1}^r)  \nonumber \\ \nonumber \\  &+&
\frac{\sin((r+1)a)\sin((N-r-1)a)}{\sin((n+r)a)\sin((n+N-r)a)}
b_n^{r+1} \nonumber \\ \nonumber \\ &+&
\frac{\sin((r-1)a)\sin((N-r+1)a)}{\sin((n+r)a)\sin((n+N-r)a)} 
b_n^{r-1},    \label{3}
\end{eqnarray}
where 
\begin{eqnarray*}
a \equiv \frac{\pi}{f+N}.
\end{eqnarray*}
We solve (\ref{3}) by inspection. Since $b_n^r$ has to satisfy the
{\it boundary conditions}, (\ref{4}), we have that
\begin{eqnarray*}
b_n^r = \sin((n+N)a)\sin(na) d_n^r
\end{eqnarray*}
is the maximal solution when $d_n^r = d= constant$,
and the eigenvalue is,
\begin{eqnarray}
\lambda = 2\cos \frac{\pi \tau}{N} = 2 \cos \frac{2\pi}{f+N}.
\end{eqnarray}
Hence, finally
\begin{eqnarray}
\tau = \frac{2N}{N+f}
\end{eqnarray}
We shall see next section that $\tau$ is the main
 critical exponent in the model.

\subsubsection[Specific heat and finite temperature
susceptibility]{Specific heat and finite temperature susceptibility} 
The expression for the impurity contribution to the free energy,
$F^i$, has always at low temperatures a term which is proportional to
$T^2$.

This contribution comes from the term proportional to 
$e^{-\frac{N}{2} \xi}$ present in $\eta_{n<f}^r$, as we discussed in the 
previous subsection.

Here, we will study contributions of the form
\begin{eqnarray}
&&\Delta F^i \propto  -\frac T{2\pi }\int_{-\infty }^\infty d\xi
\sum_{q=1}^N 
\frac{\sin \left(\pi \frac{N -q}N \right)(\alpha^q+\beta^q x_0^2)
e^{-\tau \xi} }{\cosh 
\left( \xi -\ln \frac{T_0}T\right) 
+\cos \pi \frac{N -q}N } \nonumber \\ &=&  
 -\frac T{2\pi }\left(\frac{T}{T_0}\right)^{\tau} \!\!\! \int_{-\infty
}^\infty \!\!\!\!\!\!  d\xi 
\sum_{q=1}^N 
\frac{\sin \left(\pi \frac{N -q}N \right) (\alpha^q+\beta^q x_0^2)
e^{-\tau \xi} }{\cosh 
\left( \xi\right) 
+\cos \pi \frac{N -q}N }, \nonumber \\ 
\label{F_imp_asymp}
\end{eqnarray}

\onecolumn

which might become dominant depending on the value of $\tau$.
We will consider the three cases $f>N$, $f=N$, and $f<N$ separately.

{\it Case  $f>N$:} In this case, $\tau < 1$, and 
\begin{eqnarray*}
\lim_{\xi \rightarrow -\infty} \frac{e^{-\tau \xi}}{\cosh\xi} = 0. 
\end{eqnarray*}
That means that we can make the same the approximation for the free
energy that we made when we evaluated $G_\xi
\ln(1+\eta_n^r)$. Therefore, we have
\begin{eqnarray}
F^i \sim -TS^i -T(A+B \left(\frac{H}{T} \right)^2)\left(\frac{T}{T_0}
\right)^{\tau}, 
\end{eqnarray}
with $A$ and $B$ being constants o the order of unity
and we obtain
\begin{eqnarray}
C^i \propto \left( \frac{T}{T_0} \right)^{\frac{2N}{N+f}} ,~~~
\chi^i \propto \frac{1}{T_0} \left( \frac{T}{T_0}
\right)^{\frac{N-f}{N+f}}.
\label{cv}
\end{eqnarray} 
Needless to say, when $N=2$ we recover the multichannel results. As
a matter of fact, the exponents depend only on the ratio
$\gamma=f/N$,\cite{ZarPRL,Cox,Gan,ZarVlad}.

{\it Case $f=N$:}
Since $\tau=1$ in this case, 
we cannot extend the integral in eq.~(\ref{F_imp_asymp}) to
$\xi \to -\infty$
and we have to restrict it to the interval
$[\delta,\infty)$, where $\delta$ is a finite number of the order of one. 
Making use of $
\int \frac{dz}{1+e^{2z}} = z - \frac{1}{2}\ln (1+e^{2z})$
we have for very low temperatures
\begin{eqnarray*}
\Delta F^i \! &\propto& 
\! - \frac T{2\pi } \!
\left(\frac{T}{T_0}\right) \!\! \int^{\infty 
}_{\delta-\ln \frac{T_0}{T}} \!\!\!\!\! d\xi \! 
\sum_{q=1}^N 
\frac{\left(\sin \pi \frac{N -q}N \right) \! (\alpha^q \!\! + \!\! \beta^q x_0^2)
e^{-\xi} }{\cosh 
\left( \xi\right) 
+\cos \pi \frac{N -q}N } \\ 
&\propto & -TS^i+\frac{T^2}{T_0}
\frac{\ln{T}}{T_0}(A+B\left(\frac{H}{T}\right)^2).
\end{eqnarray*}

Hence 
\begin{eqnarray}
C^i \propto -\frac{T}{T_0} \ln\frac{T}{T_0}  ,~~~
\chi^i \propto  -\frac{1}{T_0} \ln \frac{T}{T_0}.
\label{cv2}
\end{eqnarray} 

{\it Case $f<N$:}
In this region $1 < \tau < 2$.

Consider the integral
\begin{eqnarray}
\int_{\delta-\log\frac{T_0}{T}}^{\infty} \frac{e^{-(\tau-1)x} dx}{1+e^{2x}}
= \frac{1}{\tau-1} \int_{\beta
\left(\frac{T}{T_0}\right)^{\tau-1}}^{\infty}
\frac{dy}{y^2(1+y^{\frac{2}{\tau-1}})} 
\label{inte}
\end{eqnarray}
It is possible to find a primitive for $\frac{2}{\tau-1}$
integer; we have,
\begin{eqnarray*}
\int \frac{dy}{y^2(1+y^{\frac{2}{\tau-1}})} = - \frac{1}{y} + \left\{
\begin{array}{ll} \frac{1}{2n} \sum_{k=1}^{n} \cos
\frac{\pi(2n-1)(2k-1)}{2n} ~\ln (1-2y\cos \frac{\pi (2k-1)}{2n}
+y^2)  &  \\ -\frac{1}{n}\sum_{k=1} \sin \frac{\pi (2k-1)(2n-1)}{2n}
\arctan \left(\frac{y-\cos \frac{\pi (2k-1)}{2n}}{\sin \frac{\pi
(2k-1)}{2n}} \right), & \frac{2}{\tau-1} = 2n \\ \\ 
\frac{\ln(1+y)}{2n+1} + \frac{1}{2n+1} \sum_{k=1}^{n} \cos
\frac{\pi(2n)(2k-1)}{2n+1} ~\ln (1-2y\cos \frac{\pi (2k-1)}{2n+1}
+y^2)  &  \\ -\frac{2}{2n+1}\sum_{k=1} \sin \frac{\pi (2k-1)(2n)}{2n+1}
\arctan \left(\frac{y-\cos \frac{\pi (2k-1)}{2n+1}}{\sin \frac{\pi
(2k-1)}{2n+1}} \right), & \frac{2}{\tau-1} = 2n+1
\end{array} \right.
\end{eqnarray*}
As $T \rightarrow 0$, the leading terms are of the form 
\begin{eqnarray*}
\left(\frac{T}{T_0}\right)^{\tau-1} + ctn.
\end{eqnarray*}                    

\twocolumn

Hence,
\begin{eqnarray*}
\Delta F^i \propto - \left(A+B\left(\frac{H}{T}\right)^2\right)
\left(\frac{T^2}{T_0}-ctn.~ T \left(\frac{T}{T_0}\right)^{\tau}
\right) 
\end{eqnarray*}
Since $\tau > 1$, the dominant term in $\Delta F^i$ is of order
$T^2$. As for the specific heat and susceptibility, we have
\begin{eqnarray}
C^i \propto \frac{T}{T_0}-A
\left(\frac{T}{T_0}\right)^{\frac{2N}{N+f}}  ,~~~ 
\chi^i \propto \frac{1}{T_0}-B
\left(\frac{T}{T_0}\right)^{\frac{N-f}{N+f}} . 
\label{cv3}
\end{eqnarray} 
These results are valid for any $1 < \tau < 2$, as can be verified by
numerical integration of (\ref{inte}), or by the numerical solution of
the thermodynamic equations.


To summarize, there are three different kinds of behavior in the
overscreened sector, depending on the value of the ratio
$\gamma=f/N$. i) When $\gamma > 1$, both $C^i/T$ and $\chi^i$ have
power-law divergences as $T\rightarrow 0$.  The behavior is similar to
that of the multichannel Kondo model with $f>2$. Indeed, the exponents
are the same, since they depend on $\gamma$ only. ii) For $\gamma=1$
there are logarithmic divergences as in the two channel Kondo
model. iii) When $\gamma<1$, the values of $C^i/T$ and $\chi^i$ at
$T=0$ are finite. Actually, it can be deduced from the numerical
analysis that these constants are the same as in the corresponding
completely screened cases ($\mu=f$, as we will see later). However,
the fixed point has Non-Fermi liquid behavior, as can be seen from the
value of the residual entropy and from the subleading power-law
terms.

One can relate the different kinds of behavior to the shape of the
$n<f$ sector in Fig. \ref{fig1} as can be seen in the diagrams
(\ref{p1}-\ref{p3}). The square diagram corresponds to $\gamma=1$,
whereas the horizontal(vertical) one corresponds 


\subsection[Channel Anisotropy]{Channel Anisotropy}

In this section we consider briefly the case when some of the
couplings $J_m$ are different. From the study of the analogous problem
in the multichannel Kondo model \cite{aj}, we conclude that up to $f$
different energy scales will appear in the problem depending on the
pattern of symmetry breaking. The novelty  here is that there might be
a situation where $\gamma>1$ for an intermediate regime of
temperatures, whereas for very low temperatures the behavior is
characterized by an effective $\gamma$ smaller than 1.

Consider a system where the flavor symmetry is such that $p$ energy
scales $T_1<T_2<...<T_p$, are generated. Each scale $T_j$ is related
to a driving term at the level $n=m_j$ in the TBA equations
(\ref{tba}). We will assume for simplicity, that
$m_1<m_2<...<m_p=f$. If the largest flavor symmetry possible is
$SU(f)$ there will always be a driving term at the level $n=f$. 

Then, when the temperature is below any $T_j$, the thermodynamic
properties are given by (\ref{rsi}) and (\ref{cv}), where
$\gamma=f/N$ is replaced by $\gamma_{eff}=m_1/N$. As the
temperature is increased, the behavior of the system when
$T_{j-1}<T<T_{j}$ corresponds to $\gamma_{eff}=m_j/N$. Indeed, the
value to the impurity contribution to the entropy will be close to
${\cal S}^i(m_j,N)$. 

Flavor anisotropy is a relevant perturbation of the isotropic
hamiltonian. In general, the system will flow away from the fixed
point characterized by $f$, and $N$ to a new fixed point
characterized by $m_1<f$, and $N$. From (\ref{rsi}) we see that ${\cal S}^i$
is reduced in such flow (${\cal S}^i(f,N)$ is monotonous in both $f$ and
$N$). 

It is worth noticing that, once $m_1 < N$, $\chi^i$ and $C^i/T$ become
constant as $T\rightarrow 0$. 
\begin{figure}
\epsfxsize=8cm
\hskip0.5cm
\epsfbox{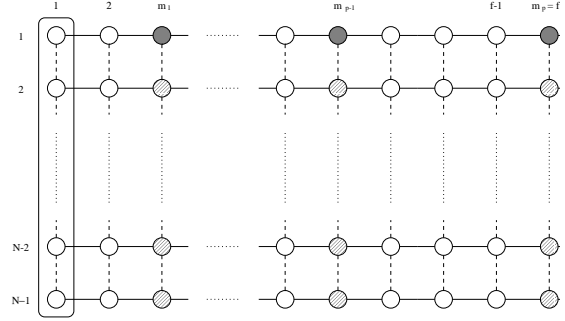}
\vskip0.7truecm  
\caption{Same as Fig. 1, but now the interaction with the impurity
breaks the flavor symmetry from $SU(f)$ down to $\prod_{j=1}^p
SU(m_j-m_{j-1})$, with $m_j^p = f$, $m_0=0$.}  
\label{fig2}
\end{figure}
The system of TBA equations are represented diagrammatically in
Fig. \ref{fig2}. Depending on the pattern of symmetry breaking, driving
terms appear at different values of $n$. There are always driving
terms for $n=f$. The properties a temperatures between two different
scales are related to the corresponding overscreened part of the
diagram. Notice that channel anisotropy may have the effect of
changing the shape in the overscreened area from something similar to
(\ref{p3}) to diagrams like (\ref{p2}), and (\ref{p3}), but not the
other way around. 

\subsection[Impurity in a higher dimensional representation]{Impurity in
a higher dimensional representation}

Finally, we study a generalization of the model in which the impurity
behavior is that of an object in a rank $\mu$ representation of
$SU(N)$. In the $SU(2)$ case it corresponds to an impurity with spin
$S$. Following Ref.~[\onlinecite{joh}] and the same formulation 
that we followed
for the fundamental representation, we find the following set of
effective Bethe Ansatz equations
\begin{eqnarray*}
e^{if p_\delta L} &=&\prod_{\gamma =1}^{M^1}\frac{\chi _\gamma
^1-1+if\frac c 2}{\chi _\gamma ^1-1-if\frac c 2}, \\
-\prod_{\beta =1}^{M^r}\frac{\chi _\gamma ^r-\chi _\beta ^r+ic}{\chi _\gamma
^r-\chi _\beta ^r-ic} &=&\prod_{t=r\pm 1}\prod_{\beta =1}^{M^t}\frac{\chi
_\gamma ^r-\chi _\beta ^t+i\frac c2}{\chi _\gamma ^r-\chi _\beta ^t-i\frac c2%
}; \\ && r=2,...,N -1, \\
-\prod_{\beta =1}^{M^1}\frac{\chi _\gamma ^1-\chi _\beta ^1+ic}{\chi _\gamma
^1-\chi _\beta ^1-ic} &=&\frac{\chi _\gamma ^1+i\mu\frac c2}{\chi _\gamma
^1-i\mu\frac c2}\prod_{\delta =1}^{N^e/f }\frac{\chi _\gamma
^1-1+if\frac c 
2}{\chi _\gamma ^1-1-if\frac c 2} \\ &\times & \prod_{\beta
=1}^{M^2}\frac{\chi _\gamma 
^1-\chi _\beta ^2+i\frac c2}{\chi _\gamma ^1-\chi _\beta ^2-i\frac c2},
\end{eqnarray*}
\begin{figure}
\epsfxsize=8cm
\hskip0.5cm
\epsfbox{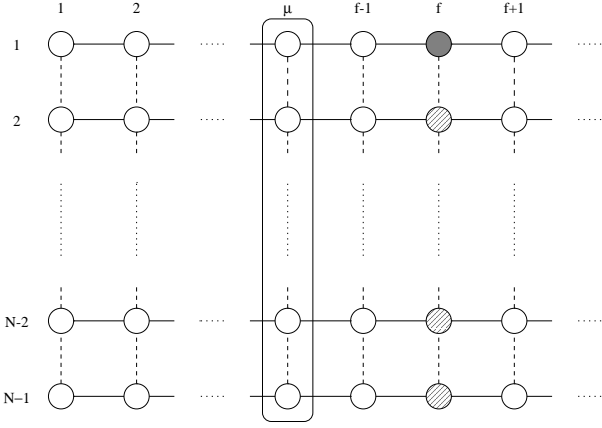}
\vskip0.7truecm
\caption{Same as Fig. 1, but now the impurity contribution to the free
energy involves the set $\eta_{\mu}^q$, where $\mu$ is the rank.}
\label{fig3}
\end{figure}
The impurity contribution to the free energy is
\begin{eqnarray*}
F^i = -T \sum_{n} \int_{-\infty}^{\infty}
B_{n,\mu}\ln(1+(\eta_n^1)^{-1}) 
\end{eqnarray*}
As in the $\mu=1$, a series of transformations allow us to rewrite the
free energy as
\begin{eqnarray*}
F^i&=&\sum_{q=1}^{N-1}\int_{-\infty }^\infty d\chi G_N ^{1,q}(\chi
)g_{\mu}^q(\chi ) \\  &-& T\sum_{q=1}^{N-1}\int_{-\infty }^\infty
d\chi G_N 
^{1,q}(\chi)\ln (1+\eta _{\mu}^q), 
\end{eqnarray*}
where the first term corresponds to the impurity contribution to the ground
state. At finite temperatures we are only interested in the second
term, which in the scaling limit can be written as 
\begin{eqnarray}
F^i=-\frac T{2\pi }\sum_{q=1}^N\int_{-\infty }^\infty \!\!\!\!\!\!  d\xi 
\frac{(\sin \pi \frac{N -q}N)\ln (1+\eta _{\mu}^q(\xi )) }{\cosh
\left( \xi -\ln \frac{T_0}T\right) 
+\cos \pi \frac{N -q}N }.
\label{freeg}
\end{eqnarray}
The evaluation of $F^i$ involves the functions $\eta_{\mu}^r$, (see
Fig. \ref{fig3}).  
The different scenarios possible are very similar to
those of the multichannel Kondo model. As long as $\mu < f$,
the impurity remains overscreened 
and the temperature exponents are the same as for the $\mu=1$ case. In
this case, the residual entropy is 
\begin{eqnarray} 
{\cal S}^i_{T=0} = \ln \frac{\prod_{r=1}^{\mu+N-1} \sin
\frac{\pi r}{f+N} }{\prod_{r=1}^{\mu} \sin\frac{\pi r}{f+N}  
\prod_{r=1}^{N-1} \sin\frac{\pi r}{f+N}}
\label{pert}
\end{eqnarray}
(Notice that when $N=2$ this reduces to the multichannel result $\ln
\frac{\sin \frac{\pi(2S+1)}{f+2}}{\sin \frac{\pi}{f+2}}$).
Furthermore, it can be easily shown that ${\cal S}^i_{\mu} 
= {\cal S}^i_{f-\mu}$. 

\begin{figure}
\epsfxsize=8cm
\hskip0.5cm
\epsfbox{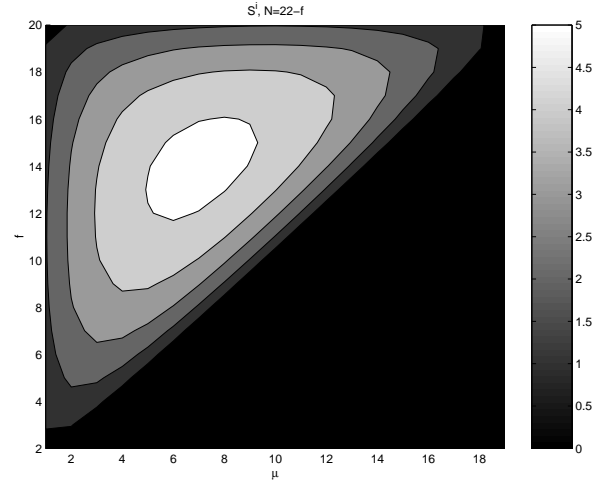}
\vskip0.7truecm  
\caption{Overscreening residual entropy, $S^i_{\mu < f}$, for an
impurity in a totally 
symmetric representation, $\mu$, of $SU(N)$, for $N+f=22$. The lower
triangular region (in black) is unphysical}
\label{entmu}
\end{figure}

We have plotted ${\cal S}^i$ in Fig. \ref{entmu}, for fixed $N+f=22$, 
and several
values of $\gamma=f/N$ and $\mu$. Only the region $\mu<f$ is physical
in the figure, since these are results for the overscreened case. 
We can see that for fixed (even) $f$, the largest value of the entropy
corresponds to $\mu=f/2$. Also, for fixed $\mu$, the entropy is the
largest around $\gamma \sim 1$, and decreases as $\gamma$ moves away
from 1.
If
$\mu=f$ the impurity becomes completely screened: $\chi^i$ and $C^i/T$
become constant, and there is no residual entropy. Finally, if
$\mu >f$, the impurity is underscreened. The dominant contribution to
the 
free energy from the spin sector is of the form
\begin{eqnarray*}
F^i = -T \ln \frac{ \prod_{r=1}^{\mu-f+N-1} \sinh
 r \frac{H}{T}}{\prod_{r=1}^{\nu-f} \sinh
r \frac{H}{T} \prod_{r=1}^{N-1} \sinh r \frac{H}{T}}
\end{eqnarray*}
When $H=0$, the residual entropy is
\begin{eqnarray*}
{\cal S}^i =  \ln \frac{(\mu-f+N-1)!}{(\mu-f)!(N-1)!} = \ln \left(
\begin{array}{c} \mu-f+N-1 \\ N-1 \end{array} \right)
\end{eqnarray*}
(For $N=2$, the residual entropy is $\ln \left(
\begin{array}{c} \mu-f+1 \\ 1 \end{array} \right) = \ln (\mu-f+1)$). 

\section[Numerical Analysis]{Numerical Analysis}

\subsection[Procedure]{Procedure}

We have solved the TBA equations by iteration, using a procedure
inspired by the work of Rajan \cite{Raj}. For the levels $n \ne f$,
which do not 
have a 
driving term (see Fig. \ref{fig3}), we have used
(\ref{easy},\ref{easy2}), as the starting 
point, since it is more convenient to use the kernel $G(\xi)$. We have
dealt with the equations that have a driving term by introducing two
sets of auxiliary functions \cite{afl},
\begin{eqnarray*}
h^r(\xi) &=& G_{\xi}(\ln(1+\eta_{f+1}^n)+\ln(1+\eta_{f-1}^n)), \\
Q^r(\xi) &=& G_{\xi}(Q^{r+1}(\xi')+Q^{r-1}(\xi')) - h^r(\xi),
\end{eqnarray*}
so that
\begin{eqnarray*}
\ln \eta_f^r = -\frac{2}{f} e^{\xi} \sin \left( \frac{\pi r}{N}
\right) + \ln(1+\eta_f^r) + Q^r.
\end{eqnarray*}

We have introduced a cutoff $A$ in the integrals involved, we have
taken $\ln(1+\eta_n^r)$ to be constant for $|\xi|>A$, and evaluated the
integrals in those intervals analytically. For $|\xi| < A$
we have replaced the integral with a sum using a Gaussian quadrature
rule \cite{NumRec}. 

The results that we present in this work correspond to zero magnetic
field, which means that $x_0 = H/T = 0$, and the functions $\eta^r_n$
depend on $\xi$ only. Thus, the task of obtaining thermodynamic
properties is greatly simplified. First of all, the impurity
contribution is given by (\ref{free})
\begin{eqnarray*}
F^i=-\frac T{2\pi }\sum_{q=1}^N\int_{-\infty }^\infty d\xi 
\frac{\left(\sin \pi \frac{N -q}N\right) \ln (1+\eta _{\mu}^q(\xi ))
}{\cosh \left( \xi -\ln \frac{T_0}T\right) 
+\cos \pi \frac{N -q}N },
\end{eqnarray*}
with $\eta_n^r$ independent of $T$. The entropy and specific heat are
obtained by taking derivatives of $F^i$ with respect to the
temperature, which 
can be done analytically when $x_0=0$, and then performing the
integration numerically.

In order to calculate the susceptibility at zero magnetic field,
$\chi^i$, we derived a second set of TBA equations for the functions 
\begin{eqnarray*}
E^r_n(\xi) \equiv \left. \frac{\partial^2 \eta_n^r(\xi)}{\partial
x_0^2} \right|_{x_0=0},
\end{eqnarray*}
following Degranges \cite{Deg}. This system is solved as the previous
one, and the magnetic susceptibility is given by
\begin{eqnarray*}
\lefteqn{\chi^i |_{x_0 = 0} = \frac{\partial^2 F^i}{\partial
x_0^2} =} && \\ &&  -\frac T{2\pi }\sum_{q=1}^N\int_{-\infty }^\infty
d\xi  
\frac{\left(\sin \pi \frac{N -q}N\right) E_{\mu}^q(\xi )
}{\cosh \left( \xi -\ln \frac{T_0}T\right) 
+\cos \pi \frac{N -q}N }.
\end{eqnarray*}

\subsection[Results]{Results}

\subsubsection[Entropy]{Entropy}

\begin{figure} 
\epsfxsize=8cm  
\hskip0.5cm\epsfbox{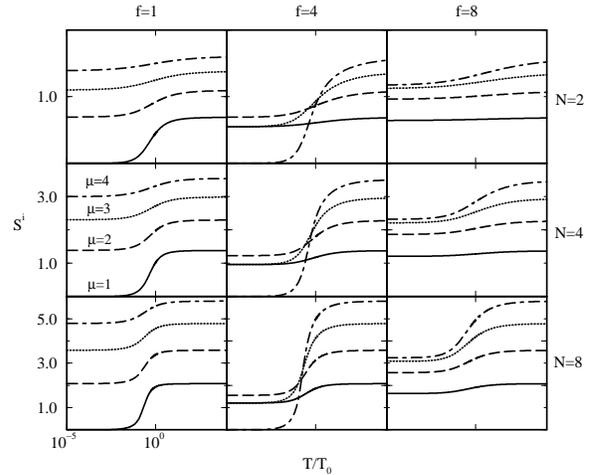}  
\vskip0.7truecm  
\caption{Impurity contribution to the Entropy as a function of $T$, for
different values of $N$, $f$, and impurity spin, $\mu$. Notice that
different scales are used for the different $N$.}
\label{fi:ent}
\end{figure}

We start by discussing the impurity contribution to
the entropy, ${\cal S}^i$. In Fig. \ref{fi:ent} we have plotted ${\cal
S}^i$ as a function of $T$, for different values of $N$, $f$, and
impurity spin $\mu$. The horizontal axis is on a logarithmic
scale. The vertical axes have different scales for the 
different $N$. The first thing to notice is the crossover around $T\sim
T_0$. For $T \gg T_0$, ${\cal S}^i$ is that of a free spin characterized by
$\mu$ and $N$. When $H=0$ 
\begin{eqnarray*}
{\cal S}^i= \ln \left(\begin{array}{c} \mu + N -1 \\ N-1 \end{array}
\right)
\end{eqnarray*}
Below the crossover region one can see the quenching of the
degrees of freedom due to the interaction in the decrease of the value
of the entropy. 
The qualitative behavior
of ${\cal S}^i$ in the region $T \ll T_0$
depends only on the relation between $\mu$ and $f$. When $f=\mu$,
there is complete screening, and ${\cal S}^i=0$, as can be seen in the
curves $\mu=f=1$, $\mu=f=4$. For $\mu > f$, the impurity is not
completely screened, and there is effectively a leftover free spin
$\mu-f$, as can be seen in Fig. 6 for
$f=1$. There, the $T \ll T_0$ entropy for $\mu<f$ corresponds to the
$T \gg T_0$ entropy for $\mu-1$. Finally, when $\mu<f$, 
overscreening takes place: even though 
there are  enough electrons to form a singlet with the
impurity, the low temperature behavior is characterized by an object
with complex internal structure, and an anomalous ${\cal
S}^i$. Such behavior can be seen in the curves $\mu=2$, $\mu=3$ for
$f=4$, and in all the curves for $f=8$. Notice that for $\mu=1$ and
$\mu = f-1$, the
curves converge to the same value, as we had already seen in the
asymptotic analysis. Furthermore, the overscreened
fixed point has an anomalous residual entropy irrespective of the
value of $N$, indicating its Non-Fermi liquid nature.

\begin{figure} 
\epsfxsize=8cm  
\hskip0.5cm\epsfbox{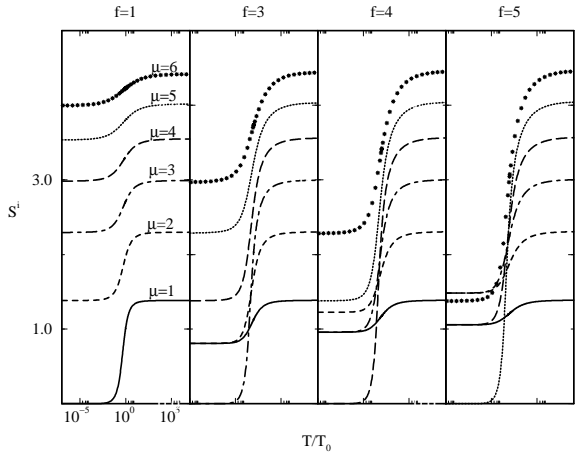}  
\vskip0.7truecm  
\caption{${\cal S}^i$ vs. $T$ for $N=4$, $f=1,3,4,5$, and $\mu=1,...,6$}
\label{fi:ent2}
\end{figure}

A more detailed picture of the behavior of ${\cal S}^i$ for $N=4$ is
displayed in Fig. \ref{fi:ent2}. It is worth noticing that in the
underscreened cases, the effective spin is $\mu-f$. Also, there might
be situations where the residual entropy of the
overscreened case is larger than that of the underscreened case. Such
is the case for $f=5$, $\mu=3,4$, as compared to $\mu=6$. 
\subsubsection[specific heat]{Specific Heat}
\begin{figure} 
\epsfxsize=8cm  
\hskip0.5cm\epsfbox{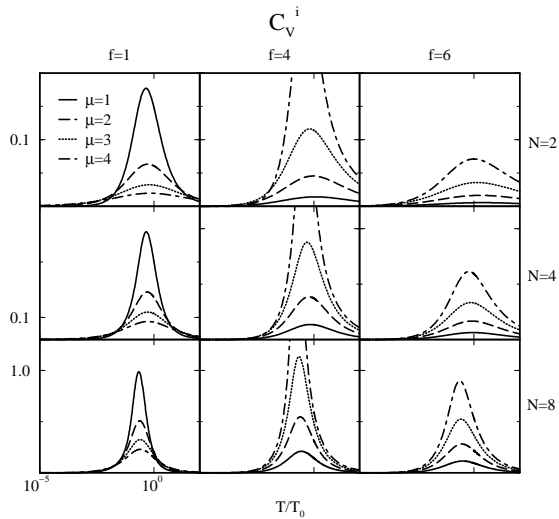}  
\vskip0.7truecm  
\caption{Impurity contribution to the specific heat, $C_V^i$, for
different values of $N$, $f$, and $\mu$}
\label{fi:cv1}
\end{figure}
Next, we compute the contribution to the specific heat. 
Results for
different values of the parameters are shown in Fig. \ref{fi:cv1}. The largest
maximum of $C_V^i$ corresponds to $\mu=f$. Also, the size of the curve 
grows with $N$.

\begin{figure} 
\epsfxsize=8cm  
\hskip0.5cm\epsfbox{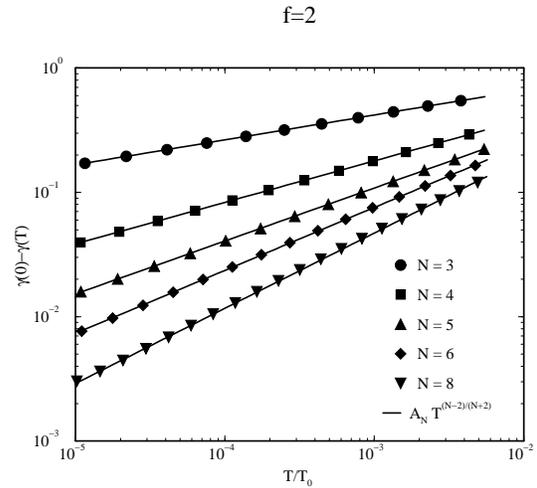}  
\vskip0.7truecm  
\caption{Subleading contribution to $\gamma(T)=C_V^i/T$, for $f=2$,
$\mu=1$, and $N>f$ vs. $T$, in a Log-Log graph. The symbols correspond
to the numerical 
calculation. The lines correspond to power-law fits with exponents
$(N-2)/(N+2)$.}
\label{fi:cv4}
\end{figure}

We have also evaluated the subleading
contribution to the linear coefficient of the specific heat, $\gamma^i =
C_V^i /T$ for $f=2\mu=2$, $N > f$, and we have plotted it in
Fig. \ref{fi:cv4}. The points fit  power-law
curves with exponents $(N-f)/(N+f)$,  derived previously (see Eqs.~(\ref{cv}), 
(\ref{cv2}), and (\ref{cv3})). 
. This
is another clear indication that for $N > f$, the overscreened cases
are not Fermi liquid fixed points.

\subsubsection[Susceptibility]{Magnetic Susceptibility}

\begin{figure} 
\epsfxsize=8cm  
\hskip0.5cm\epsfbox{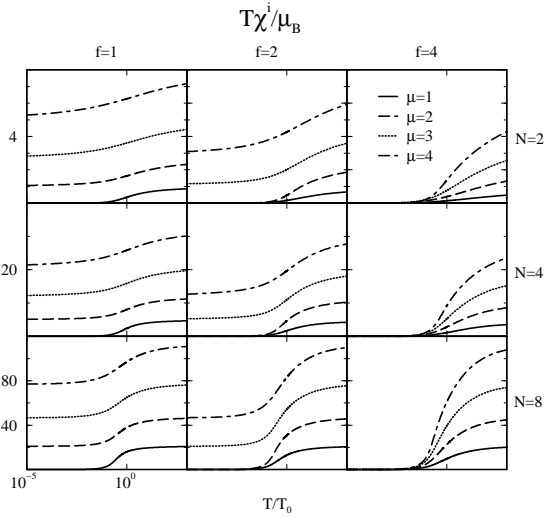}  
\vskip0.7truecm  
\caption{$T\chi^i$ vs. T, for different values of $f$, $N$, and $\mu$}
\label{fi:chi1}
\end{figure}

Next, we have studied $T\chi^i$ for different values of the
parameters, and plotted the results on Fig. \ref{fi:chi1}. As with the
entropy, the qualitative behavior 
depends only on the values of $\mu$ and $f$. The 
difference in behavior between underscreened and 
overscreened cases becomes more clear here: whereas the magnetic
moment is partially quenched 
in the former case, the overscreened case is characterized by a
totally quenched moment, even though there is a non-zero residual
entropy. This can be seen in  Fig. \ref{fi:chi1} for the curves with 
$f=1$ and $f=4$. 

\begin{figure} 
\epsfxsize=8cm  
\hskip0.5cm\epsfbox{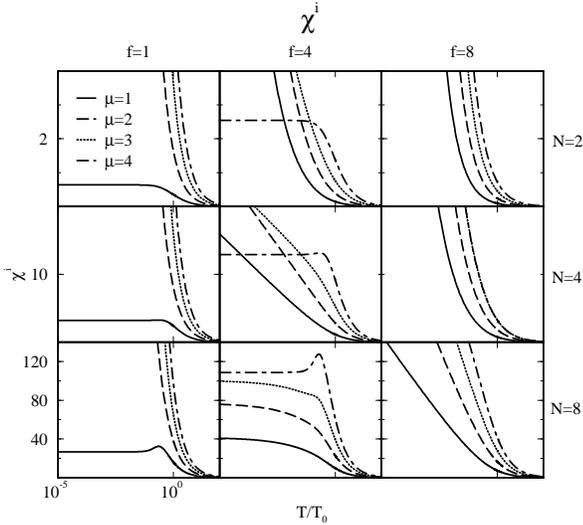}  
\vskip0.7truecm  
\caption{Impurity contribution to the magnetic susceptibility,
$\chi^i$ vs. $T$, for different values of $f$, $N$, and $\mu$.}
\label{fi:chi2}
\end{figure}

\begin{figure} 
\epsfxsize=8cm  
\hskip0.5cm\epsfbox{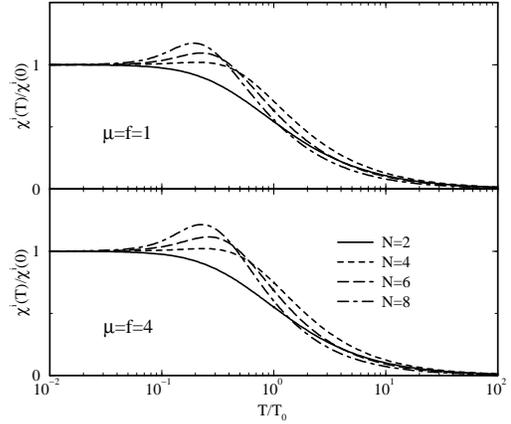}  
\vskip0.7truecm  
\caption{$\chi^i(T)/\chi^i(0)$ vs. $T$ for different values of $N$, in
two completely screened cases, $\mu=f=1$, and $\mu=f=4$.}
\label{fi:chi3}
\end{figure}

The magnetic susceptibility is plotted in Fig. \ref{fi:chi2}. The curves
with $f=\mu$ have a constant $\chi^i$ at low-T and for $N > 2$ they
have a maxima near $T \sim T_0$. This is a special feature of the
completely screened case.  We see that in the
overscreened case with $N > f$, the susceptibility tends to a finite
value as $T \rightarrow 0$, while it diverges when $N \le f$. When
$N=f$, the divergence is logarithmic, whereas it is power-law for
$N<f$, with an exponent $\beta > -1$. The largest divergence
corresponds to the underscreened case with $1/T$ behavior. All these
results coincide with those of the previous analytic study,
Eqs.~(\ref{cv}), (\ref{cv2}), and (\ref{cv3}).   
In Fig. \ref{fi:chi3} we
show $\chi^i$ for different values of $N$ for the cases $\mu=f=1$ and
$\mu=f=4$. We have rescaled the curves dividing by $\chi^i(0)$. It is
quite apparent that the behavior is the same in both cases.
\begin{figure} 
\epsfxsize=8cm  
\hskip0.5cm\epsfbox{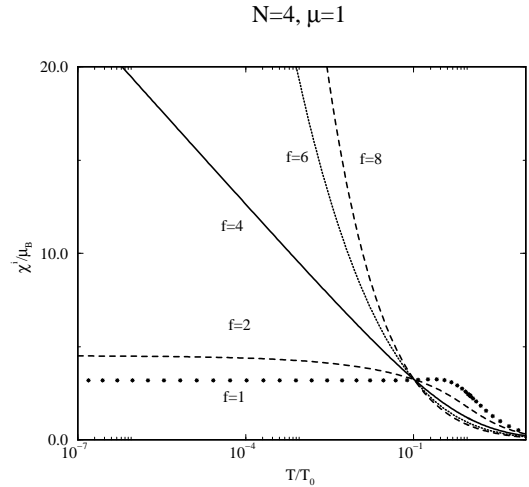}  
\vskip0.7truecm  
\caption{$\chi^i$ vs. $T$ for $N=4$, $\mu=1$, and $f=1,2,4,6,8$.}
\label{fi:chi4}
\end{figure}

\begin{figure} 
\epsfxsize=8cm  
\hskip0.5cm\epsfbox{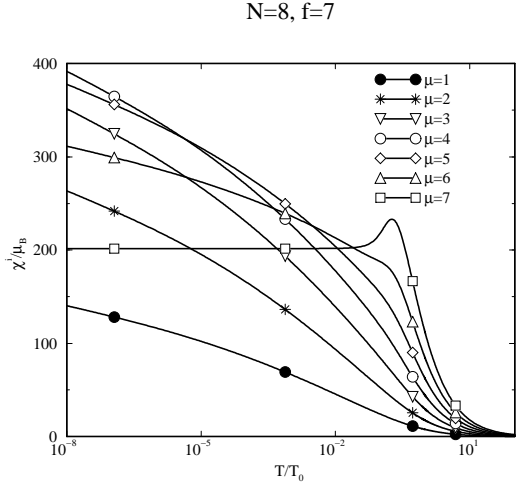}  
\vskip0.7truecm  
\caption{$\chi^i$ vs. $T$ for $N=8$, $f=7$, and $\mu=1,...7$.}
\label{fi:chi8}
\end{figure}

In Fig. \ref{fi:chi8} we consider $N=8$,
$f=7$ and $\mu=1,...,7$. Even though $\chi^i(0)$ is finite, it is
clear that the behavior of the overscreened case is quite different
from that of the completely screened case, and that the subleading terms 
have an important contribution below the crossover temperature. 

\begin{figure} 
\epsfxsize=8cm  
\hskip0.5cm\epsfbox{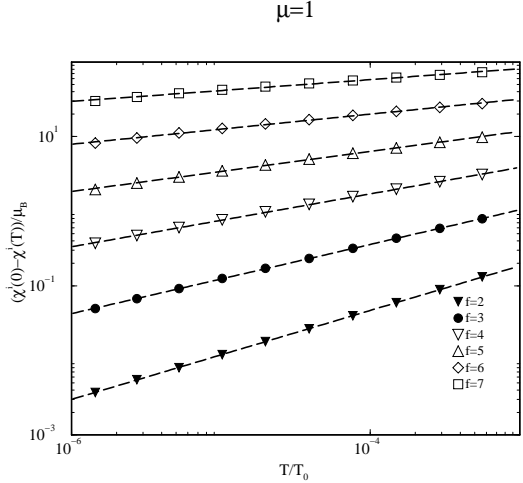}  
\vskip0.7truecm  
\caption{Subleading T dependence of $\chi^i$ for $N=8$, $f=2,...,7$,
$\mu=1$. The lines correspond to power law fits.}
\label{fi:chi9}
\end{figure}

We can also see the power-law behavior of the subleading term of
$\chi^i$ 
in Fig. \ref{fi:chi9}. The values for the exponents agree with the
values obtained analytically, i.e. $(N-f)/(N+f)$.

\subsubsection{Wilson Ratio}

\begin{figure} 
\epsfxsize=8cm  
\hskip0.5cm\epsfbox{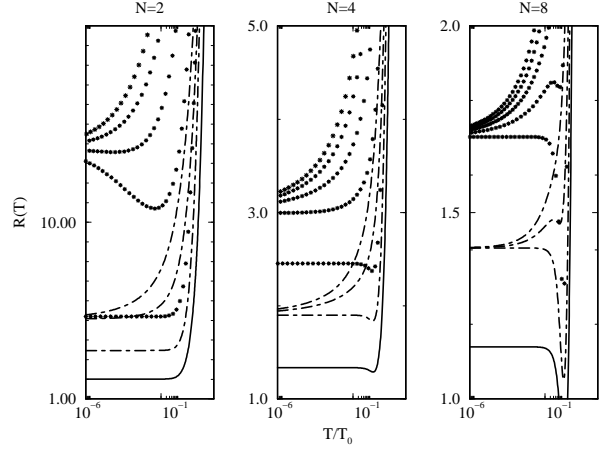}  
\vskip0.7truecm  
\caption{$R(T)$ as a function of temperature for $N=2,4,8$,
$\mu=1,...,f$ and $f=1,3,5$.}
\label{fi:r1}
\end{figure}

We have calculated the Wilson ratio, defined as
\begin{eqnarray}
R \equiv \frac{\pi^2 k_B^2}{(N^2-1)\mu_B^2} \frac{T \chi^i}{C_V^i}
\end{eqnarray}
The quantity $R$ has a well-defined meaning only for $T=0$. However,
in Fig. \ref{fi:r1} we have plotted the quantity $R(T)$, to show the
difference between the $N < f$ and the $N > f$ sectors. In the former case,
($N=2$, $f>1$), 
the value for the overscreened case is much larger than the value of
the completely screened case (notice the difference in vertical
scales), whereas in the latter case  ($N=8$), the 
curves 
converge to the completely screened value. Notice
that for $N=4$ there is a change in behavior as we go from $f < N$ to
$f > N$. 

\begin{figure} 
\epsfxsize=8cm  
\hskip0.5cm\epsfbox{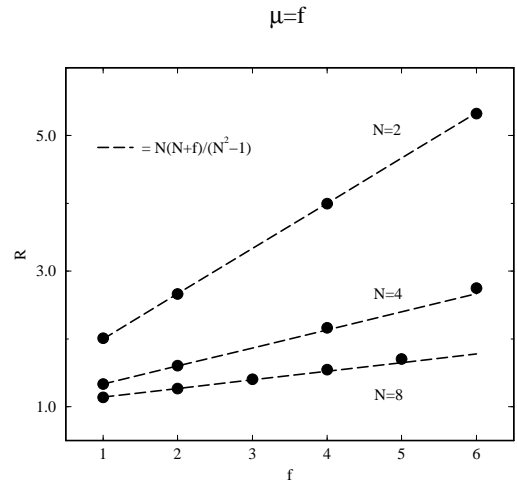}  
\vskip0.7truecm  
\caption{Wilson Ratio R for $\mu=f$. The points have been obtained
from the numerical solution. The lines correspond to fits with
the function $N(N+f)/(N^2-1)$.}
\label{fi:r2}
\end{figure}

\begin{figure} 
\epsfxsize=8cm  
\hskip0.5cm\epsfbox{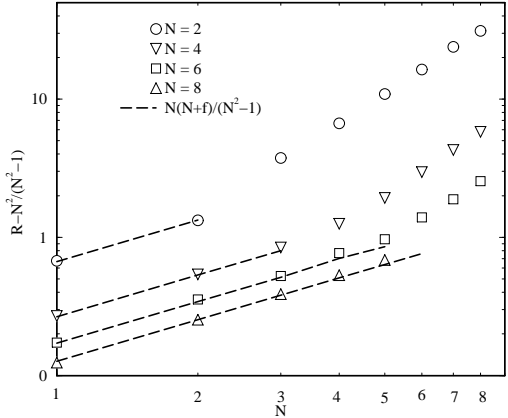}  
\vskip0.7truecm  
\caption{Wilson Ratio $R$ for $\mu < f$. See the caption in the
previous figure.}
\label{fi:r3}
\end{figure}

Next we plot the values of $R$ for completely screened cases, 
$\mu=f$ (Fig. \ref{fi:r2}). We see that
the values obtained fit the function
\begin{eqnarray}
R = \frac{N(N+f)}{N^2-1}.
\label{reo}
\end{eqnarray}
We have also obtained values of $R$ for the overscreened case
($\mu<f$), in 
Fig. \ref{fi:r3}. There are clear differences between the $f<N$ and the
$f>N$ cases, as we have already pointed out. For $f<N$, the value of
$R$ coincides with the value for $\mu=f$, and agree with
(\ref{reo}). The dominant contribution to
$R$ comes from the constant terms in $\chi^i$ and $C_V^i/T$. For
$f>N$, $R$ contains mainly the coefficients of the divergent parts,
and have a different functional behavior. 
\subsubsection[Channel Anisotropy]{Channel Anisotropy}

\begin{figure} 
\epsfxsize=8cm  
\hskip0.5cm\epsfbox{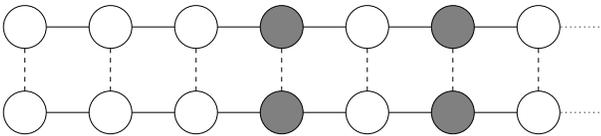}  
\vskip0.7truecm  
\caption{Diagrammatic representation of the TBA equations for a system
with $N=3$, and flavor symmetry $SU(6)$ broken down to $SU(4)\times
SU(2)$. Two driving terms appear at $n=4,6$}
\label{fi:ani}
\end{figure}

\begin{figure} 
\epsfxsize=8cm  
\hskip0.5cm\epsfbox{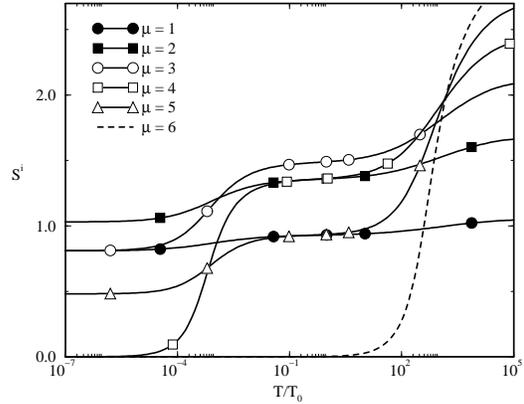}  
\vskip0.7truecm  
\caption{Entropy vs. T, for $\mu=1,...,6$}
\label{fi:anient}
\end{figure}

We end the discussion of the numerical results by showing an example
of channel anisotropy. We have taken the case $N=3$, with original
flavor symmetry $SU(6)$, broken down to $SU(4)\times SU(2)$
(Fig. \ref{fi:ani}) Two scales appear in this problem: $T_6$, and
$T_4$. Accordingly, the TBA equations have driving terms at
$n=4,6$. 
We have chosen very small anisotropy, $T_4/T_6 = 10^{-6}$.  
In Fig. \ref{fi:anient} we have plotted the entropy for different values
of $\mu$. There are three different regions: When $T>>T_6$, the
impurity behaves like a free moment. Around $T\sim T_6$ there is a
crossover to an overscreened region (when $\mu<6$), characterized by a
$SU(6)$ flavor symmetry. The pairs of curves $(\mu,6-\mu)$ merge. 
Notice also that $\mu=6$ is completely
screened. There is a second crossover around $T \sim T_4$ to
a region characterized by $SU(4)$ flavor symmetry for $\mu < 4$ and by
$SU(2)$ flavor symmetry for $4 < \mu < 6$. Only the curves for $\mu=1$
and $\mu=4-1=3$ coincide now. Also, the value of the residual entropy
for $\mu=1$, is that of an effective $\mu=1$ in a $SU(3)\times SU(2)$
model. The entropy for $\mu=4$ goes to 0 with $T$ since the system
becomes screened for $T \ll T_4$.  
\begin{figure} 
\epsfxsize=8cm  
\hskip0.5cm\epsfbox{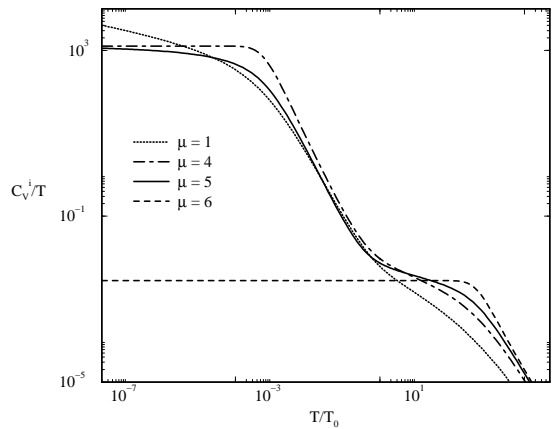}  
\vskip0.7truecm  
\caption{$C_V^i/T$ vs. T, for $\mu=1,4,5,6$.}
\label{fi:anicvt}
\end{figure}
Finally, we have plotted $C_V^i/T$ for several $\mu$ in
Fig. \ref{fi:anicvt}. When $T_4 << T << T_6$, the behavior of the
system is characterized by $f=6$: increasing value of $C_V^i/T$ for
$\mu<6$, constant behavior for the screened case, $\mu=6$. Near $T_4$
the curves for $\mu<6$ have a similar behavior as those for free
moments (however, such behavior is not found in the curves for the
magnetic susceptibility). We can see three different behaviors in the
region $T \ll T_4$. For $\mu=1$ the curve diverges with a power law,
as one would expect for $f_{eff}=4 > N=3$. The curve for $\mu=4$ is
flat, 
since the impurity is screened. Finally, for $\mu=5$, the curve
increases slowly, converging to a constant value, as one would expect
from an effective $f_{eff} = 2 < N=3$.

\section{Physical realizations of the $SU(N)\times SU(f)$ model}

In the present section we shall concentrate on the possible physical 
realizations and applications of the $SU(N)\times SU(f)$ 
Coqblin-Schrieffer model. 

In Subsections.~\ref{ss:fixpoint}, \ref{ss:stability}, and 
\ref{ss:physicalq} we 
analyze the so called N-level system (NLS) model,
a generalization of the two-level system model,\cite{ZawVlad} by means 
of a systematic $1/f$ expansion. In this context the flavor degeneracy
is associated with the physical spin of the electron. 
First we establish a mapping of the NLS model to the multichannel
Coqblin-Schrieffer model (MCCS model) by analyzing 
the low-energy fixed point of its scaling equations. While our 
procedure gives a systematic expansion only for the case $f>N$ we 
shall argue that the same mapping should apply  for the cases $f\le N$. 

In the limit $f>N$ we are able to determine the full operator content of
the fixed point. This enables us to   calculate the scaling of the 
different physical quantities at low temperatures in 
Subsection~\ref{ss:physicalq}. As we shall see there are 
some subtle differences between the two models and while most of the physical 
quantities show the same dependence, the scaling of the specific heat
may be different. The origin of these differences will be 
discussed in detail. 

Finally, based on the results of 
Subsections~\ref{ss:fixpoint}-\ref{ss:physicalq}, we discuss in 
Subsection~\ref{ss:physreal}  some 
physical systems providing possible candidates for the realization of 
the MCCS-model.

\subsection{The N-level system model and its low-energy fixed point}
\label{ss:fixpoint}

The NLS model has been constructed as a generalization of the
two-level system model\cite{ZawVlad,Kondo} to describe the tunneling of
a heavy particle among $N$ not necessarily equivalent positions 
labeled by $a=\{1,..,N\}$,
and strongly coupled to the conduction electrons. At low
temperatures the motion of the heavy particle can be described
by the effective Hamiltonian
\begin{equation}
H_{\rm hp} = \sum_{a,b=1}^N \chi^+_a \Delta^{ab} \chi_b \; , 
\end{equation}  
where $\chi^+_a$ creates a pseudofermion\cite{Abrikosov} 
corresponding to the heavy particle site $a$ and $\Delta^{ab}$ 
is  the tunneling amplitude between positions $a$ and $b$.  
If no external stress is present  then the diagonal part
 of $\Delta^{ab}$ vanishes: 
$\Delta^{aa}=0$ ($a=1,..,N$), when the 
$N$ positions are  equivalent due to the symmetry of the
NLS. The electronic part of the Hamiltonian
and the coupling of the heavy particle to the conduction 
electrons take the general form:
\begin{eqnarray} 
H_{\rm el}&=& \sum_{\epsilon n m} \epsilon \;
c^+_{\epsilon n m} c_{\epsilon   n   m} \;,  \nonumber \\
H_{\rm e-hp} &=& \sum_{\scriptstyle a,b,n,n^\prime  
\atop\scriptstyle 
\epsilon, \epsilon^\prime,m} c^+_{\epsilon n m} \chi^+_a
V_{nn^\prime}^{ab} \chi_b  c_{\epsilon^\prime   n^\prime   m}  
\; , 
\end{eqnarray}  
where the operators $c^+_{\epsilon n m}$ create  conduction
electrons with energy $\epsilon$, orbital quantum number\cite{ZawVlad} ($\sim$
angular momentum) $n=1,2,..,\infty$, and spin $m$. For the sake of simplicity
the electronic density of states $\varrho(\epsilon)$ is assumed to
be constant  $\varrho_0$,  between the high- and low-energy cutoffs, $D$ and
$-D$, independently of the flavor and orbital quantum numbers. While in the 
physical case only $m=\pm$ is possible corresponding to the two
different spin directions, for technical reasons in the following
we assume that the electron spin $m$ can take $f$
different values: $m=1,..,f$. 

This model has a structure similar to Eq.~(\ref{ham}) but there
are some important differences. The 'spin index' $a'=1,..,N$ of
the impurity in Eq.~(\ref{ham}) is now replaced by the 'site index' 
$a=1,..,N$ of the heavy particle. Moreover, in the NLS
case the orbital index $n$ (replacing the 'spin index' $a=1,..,N$ of
the conduction electrons in the $SU(N)\times SU(f)$ model) now
ranges from one to infinity since the conduction electrons
may have any orbital momentum. Furthermore, the couplings are {\it
highly anisotropic} in  orbital indices and no 
 $SU(N)$ symmetry is present at this level. Finally,
in the NLS model the scattering is diagonal in the real spin
index $m$ which plays now the same role as the flavor in
Eq.~(\ref{ham}). 

The diagonal couplings, $V_{nn^\prime}^{aa}$ describe simple potential
scattering of the conduction electrons by the heavy particle sitting
in position $a$. On the other hand, the off-diagonal matrix elements,
$V_{nn^\prime}^{ab}$ with $a\not=b$ correspond to the so called "assisted tunneling"
processes. Here the heavy particle is tunneling from one site to 
another while a conduction electron is scattered by it. The combination of 
these two processes leads ultimately to the generation of an orbital
Kondo effect and a strongly correlated ground state.\cite{ZawVlad,ZarPRL}

In the following we shall carry out a large $f$ analysis to determine the 
low-energy fixed point of the NLS model. While our procedure is strictly 
valid only in the case $f>N$, in the end of the subsection we
shall argue that our results are very general and they should apply
even for the cases $f\le N$.
 
To carry out a $1/f$ analysis of the NLS model, as a next step,
we construct the next to leading logarithmic scaling equations
using a generalized multiplicative renormalization group
technique.\cite{ZarPRL} As discussed in
Refs.~\onlinecite{Nozieres,Gan,ZarVlad,Muramatsu}, the leading 
logarithmic equations give the leading term in a systematic
$1/f$ expansion and  become exact in the $f\to \infty$
limit. In the multiplicative renormalization group method one
exploits the existence of a non-trivial transformation in the
space of the Hamiltonians,  $D\to D\prime$, $V^{ab}_{nn^\prime} \to
{V^\prime}^{ab}_{nn^\prime}$, and $\Delta^{ab}\to 
{\Delta^\prime}^{ab}$ that leaves the pseudofermion Green's
function ${\cal G}^{ab}$ and the pseudofermion-conduction electron vertex
function $\Gamma^{ab}_{nn^\prime}(\omega)$ invariant:
\begin{eqnarray} {\cal G}(\omega, V^\prime,  
\Delta^\prime,  D^\prime) & = & A \; {\cal 
G}(\omega, V,\Delta, D)\; A^+ \; , \nonumber \\  
\Gamma(\omega, V^\prime, \Delta^\prime, D^\prime) & = &  
[ A^+]^{-1}  \Gamma(\omega, V,\Delta,D) \; A^{-1} \; .
\label{eq:mrg}  
\end{eqnarray}  
In these equations $A$ denotes an $N\times N$
matrix independent of the energy variables, $\omega$ and $T$,
and acting in the site indices: $A=A^{ab}(V^\prime, \Delta^\prime,
D^\prime/D)$. By means of the transformation Eq.~(\ref{eq:mrg})
one can generate effective Hamiltonians that describe the system's
behavior below the energy scale $D'$. The generated effective Hamiltonians 
usually (in a renormalizable theory) turn out to be  much simpler than the
original one.
\begin{figure} 
\epsfxsize=8cm  
\hskip0.5cm\epsfbox{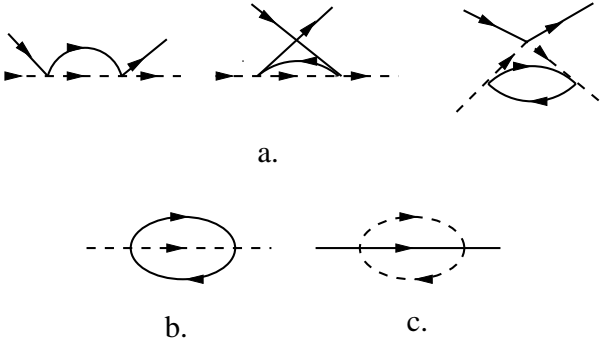}  
\vskip0.7truecm  
\caption{\label{fig:nll} The leading logarithmic  vertex 
and self-energy diagrams 
generating the next to leading logarithmic scaling equations.
Continuous and dashed lines represent the conduction electron
and pseudofermion Green's functions.} 
\end{figure}
To make use of the invariance property Eq.~(\ref{eq:mrg}) one
first has to construct the lowest order vertex and 
pseudofermion self-energy corrections\cite{ZawVlad} arising from the
diagrams in Fig.~\ref{fig:nll}.a and b,
\begin{eqnarray}
\Sigma^{ab} & =& - f \ln {D\over \omega} \;   
(\delta^{ab}\;\omega \;{\rm tr}\{ \v^{cd} \v^{dc} \} - {\rm  
tr}\{ \v^{ac} \Delta^{cd}  
\v^{db} \}) \nonumber \\ 
\varrho_0 {\tens \Gamma}^{ab} & = & \v^{ab} - \ln{D \over \omega}  
\;  
\Big( [\v^{ac},\v^{cb}] - f\; {\rm tr} \{ \v^{ac}  \v^{db} \}\v^{cd}  
\Big), 
\label{eq:vertex} 
\end{eqnarray} 
where $\varrho_0$ is the density of states at the Fermi level,
and a matrix notation has been introduced for the
dimensionless couplings $\varrho_0\;V^{ab}_{nn^\prime}
\to \v^{ab}$. The symbol $[\phantom{m},\phantom{m}]$ stands
for the commutator, the trace operator ${\rm tr}\{...\}$ is
acting in the electronic indices, and a summation must be carried out 
over repeated indices. Then, substituting Eq.~(\ref{eq:vertex}) into
Eq.~(\ref{eq:mrg}) and reducing the bandwidth $D$ by an
infinitesimal amount one can deduce the infinitesimal
renormalization group transformations for the couplings:
\begin{eqnarray} 
{d\Delta^{ab}\over dx} & = & -{1\over 2} f \Big[ {\rm  
tr}\{ \v^{ac} \v^{cd} \} \Delta^{db} + \Delta^{ac} 
{\rm tr}\{ \v^{cd} \v^{db} \}  \Big. \nonumber  \\ &&- 2
{\rm tr}\{ \v^{ac} \Delta^{cd}    
\v^{db} \}  \Big]   \label{eq:scaling1} \\
{d\v^{ab}\over dx} & = & - [\v^{ac},\v^{cb}] + {1 \over 2}  
f  \Big(2 {\rm tr} \{ \v^{ac}  \v^{db} \}\v^{cd} \nonumber \\
&& - {\rm tr}\{  \v^{ac}  
\v^{cd} \}  \v^{db} - \v^{ac} {\rm tr}\{ \v^{cd} \v^{db} \} \Big),
\label{eq:scaling2}  
\end{eqnarray}  
where the dimensionless scaling variable $x=\ln{(D_0/D)}$ has
been introduced, $D_0$ being the initial (real) bandwidth cutoff
of the model. 

These scaling equations have to be solved with the
boundary condition that the couplings are equal to their bare
values at $x=0$, and they loose their validity if the reduced
bandwidth $D$ becomes smaller than any small-energy scale
present, $T$, $\omega$, $\Delta$. Note that, up to the next to
leading logarithmic order, the splittings $\Delta^{ab}$ do not occur
in Eq.~(\ref{eq:scaling2}) explicitly, and they provide only a
low-energy cutoff for the scaling. To be explicit, there is an
energy scale, $D^*=T^*$ that we call the {\it freezing temperature},
where the renormalized splitting becomes of the same order of
magnitude as the reduced bandwidth: $\Delta^{ab}(D^*)\sim D^*$. 
Below this energy scale the orbital motion of the NLS is usually
frozen out (see the discussion in the end of this Section), and
the couplings may be replaced by their values at $T^*$.

\begin{figure} 
\epsfxsize=7cm  
\hskip0.5cm\epsfbox{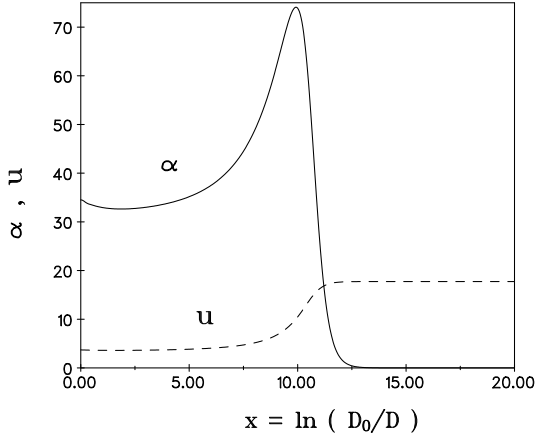}  
\vskip0.7truecm  
\caption{\label{fig:U's} Scaling of the norm of the  
dimensionless  
couplings, $u=\sum ||v^{ab}||$ (dashed line), and of the  
algebra  coefficient $\alpha$ (continuous line) for a 6-state 
system  with $f=2$.} 
\end{figure}

For the moment let us forget about Eq.~(\ref{eq:scaling1}) and
 concentrate on the scaling of the $\v^{ab}$'s,
Eq.~(\ref{eq:scaling2}). This equation cannot be solved
generally, but one can convince oneself very easily that if the
'assisted tunneling' matrix elements $\v^{ab}$ ($a\not= b$) do
not vanish then the electron-NLS couplings start to increase
and lead to a Kondo effect.\cite{ZawVlad,ZarVlad} The scaling of
the norm of the 
couplings, $\sum_{a,b}||\v^{ab}||$ is shown in
Fig.~\ref{fig:U's} for a symmetrical six-state system, where the
coupling constants have been estimated using similar methods
as in Ref.~\onlinecite{ZawVlad}. As one can see, a Kondo effect
occurs around the Kondo scale $T_K \sim D_0\;e^{-x_c}\sim 10K$,
where $x_c= \ln{(D_0/T_K)}$ denotes the value of the scaling
parameter at which the crossover from weak to strong coupling
occurs.  Our numerical investigations for various model
parameters and different values of $N$ show that the structure
of the stable low-temperature fixed point the couplings scale to is {\it 
independent} of the initial couplings and only depends on the
value of $N$ as long as no special symmetry is assumed for
the $\v^{ab}$'s. 

In what follows we shall show that this stable
low-energy fixed point of Eq.~(\ref{eq:scaling2}) has 
 the structure of the defining representation of the
$SU(N)$ Lie algebra. To be precise we first observe that 
the operators ${\cal O}^a\sim\delta^{ab}\sum_c \v^{cc}$  are invariant 
under scaling. Therefore the
$\v^{ab}$'s can be divided into two parts, ${\tilde\v }^{ab}$
and $\M^{ab}$ where $\sum_a {\tilde\v }^{aa}= 0$ and $\M^{ab}$ 
is built up from the previously
mentioned constants of motions, ${\cal O}^a$. Then as we shall see, at
the stable fixed points of Eq.~(\ref{eq:scaling2}) the
${\tilde\v}^{ab}$'s can be written as
\begin{equation} 
({\tilde \v}^{ab})_{\rm fp} = {1\over f} \left(\matrix{\L^{ab}
& 0 \cr 0 & 0  \cr}\right)\; , 
\label{eq:fp} 
\end{equation} 
where the $\L^{ab}$'s satisfy the standard $SU(N)$ Lie algebra,
\begin{equation} 
[\L^{ab},\L^{cd}] = \delta^{ad} \L^{cb} - \delta^{cb}  
\L^{ad}\;. 
\label{eq:Lie}
\end{equation} 
and are unitary equivalent to the defining representation:
$L^{ab}_{nn^\prime} \sim \delta^a_{n^\prime} \delta^b_n -
{1\over f} \delta^{ab} \delta_{nn^\prime}$. This statement is
also demonstrated in Fig.~\ref{fig:U's} 
where the scaling of the 'algebra  coefficient' 
$\alpha= \sum_{a,b,c,d} || f^2 [{\tilde \v}^{ab},{\tilde 
\v}^{cd}] - f  
\delta^{ad} {\tilde \v}^{cb} + f \delta^{cb} {\tilde  
\v}^{ad} || $ is shown, measuring how well the fixed point  algebra
(\ref{eq:Lie}) is satisfied. As one can see in Fig.~\ref{fig:U's}
for $D\ll T_K$ the algebra coefficient $\alpha$ vanishes and therefore,
in an appropriate basis,  the ${\tilde \v}^{ab}$'s really simplify to 
the form in Eq.~(\ref{eq:fp}).
Eq.~(\ref{eq:fp}) means that {\it at the fixed point}
${\tilde  \v}^{ab}$ is given by Eq.~(\ref{eq:fp}) and 
apart from some potential scattering term the fixed point
effective interaction can be written as 
\begin{equation}
H_{\rm eff} = V_0 \sum_{\scriptstyle a,b  
\atop\scriptstyle 
\epsilon, \epsilon^\prime,m} \chi^+_a 
c^+_{\epsilon b m} c_{\epsilon^\prime a  m} \chi_b  \; , 
\label{eq:H_eff}
\end{equation}  
which is the same as the interaction term in Eq.~(\ref{ham}). Note
that while the initial 
model was very asymmetrical in the orbital space, at the
fixed point {\it only $N$ conduction electron angular momentum
channels}  are coupled to the NLS and the fixed
point effective Hamiltonian shows already an additional $SU(N)$
symmetry in the orbital sector as well. These statements are not
true away from the fixed point where various kinds of 
irrelevant operators couple the NLS to the electrons and
coupling to the other orbital channels is also relevant. 
The effective Hamiltonian is completely symmetrical in the NLS
site index, which means that, e.g., the amplitude of assisted
tunneling from site 1 to 6 in Fig.~\ref{fig:U's} is the same as
the nearest neighbor assisted tunneling amplitude from site 1
to 2, despite of their different geometrical position. 

As it will become obvious in the next subsection, the analysis above
is based on the possibility of a systematic $N/f$ expansion. Therefore,
it is strictly valid in the $f>N$ case. However, one has several arguments
that the effective Hamiltonian Eq.~(\ref{eq:H_eff}) is also adequate for
the $N\le f$ cases. First of all, in the case $N=2$ corresponding to
the simpler case of  multichannel Kondo model, it is well-known that 
for $f=1$ and $f=2$ the spin anisotropy of the couplings is {\it irrelevant}
 around the fixed point,\cite{Nozieres,PangCox} which has the 
same $SU(2)$ structure as Eq.~(\ref{eq:fp}). Furthermore, for 
$f=1$ but {\it arbitrary} $N$ one can easily prove following similar lines
as Nozi\`eres and Blandin\cite{Nozieres} that the isotropic 
fixed point, Eq.~(\ref{eq:H_eff}), is stable against spin (orbital)
anisotropy. These observations together with our results for
the $f>N$ case make it highly improbable that for $2\le f\le N$
the NLS model would have a stable fixed point different from the one 
discussed above.

\subsection{Stability analysis of the $SU(N)\times SU(f)$ fixed point of the
NLS model in the large $f$ limit}
\label{ss:stability}

The statement that ${\tilde \v}^{ab}$ is a fixed point of
Eq.~(\ref{eq:scaling2}) is trivial. However, we also want to prove the
stability of this fixed point analytically and find  the irrelevant
operators (determining the low-energy behavior of the model)
around it. To this end we write the deviations from the fixed
point in the form:
\begin{equation}
 \dv^{ab} =
\left(\begin{array}{cc}
{\tens \varrho}^{ab} & \t^{ab} \\
(\t^{ba})^+ & {\tens \mu}^{ab}
\end{array}\right)
\;, \label{eq:deviation}
\end{equation}
where the couplings ${\tens \varrho}^{ab}$, $\t^{ab}$, and
${\tens \mu}^{ab}$ are $N\times N$, $N\times \infty$, and
$\infty \times \infty$ matrices, respectively. 
Substituting this expression into Eq.~(\ref{eq:scaling2}) one
obtains the following linearized decoupled  scaling equations:
\begin{eqnarray} 
{d {\tens \mu}^{ab}\over dx} &=& {1\over f} \left(
\delta^{ab}{\tens \mu}^{dd} - N {\tens \mu}^{ab}\right)\;,
\label{eq:mscaling} \\ 
 {d{\tens \varrho}^{ab}\over dx} & =& - {1 \over f} 
\left([\L^{ad},{\tens\varrho}^{db}] + 
[{\tens\varrho}^{ad},\L^{db}] \right) \nonumber \\
&+& {1\over 2f} \Bigl\{2\delta^{ab}{\tens\varrho}^{dd} +  2 \L^{cd}  {\rm 
tr} \{{\tens\varrho}^{ac}  
\L^{db} + \L^{ac}{\tens\varrho}^{db}\} 
\nonumber \\ 
&-& 2N\;{\tens\varrho}^{ab}
-\L^{ac}{\rm tr}\{ {\tens\varrho}^{cd}\L^{db} + \L^{cd} {\tens\varrho}^{db}\} 
\nonumber \\ &-& {\rm 
tr}\{{\tens\varrho}^{ac}\L^{cd}
+\L^{ac}{\tens\varrho}^{cd}\}\L^{db}\Bigr\}\label{eq:roscaling} \\
{dt^{ab}\over dx} &=& -{1 \over f} 
\left(\L^{ad}t^{db} - \L^{db}t^{ad}\right) + {1\over f}
\left(\delta^{ab}t^{dd} -  N t^{ab}\right)
 \;. \nonumber \\ && \label{eq:tscaling}
\end{eqnarray}
The solution of Eq.~(\ref{eq:mscaling}) is trivial, since the
operator ${\tens \mu}^{ab}$ can be decomposed as
${\tens \mu}^{ab}= ({\tens \mu}^{ab} - \delta^{ab}{1\over N}
{\tens \mu}^{cc}) + ( \delta^{ab}{1\over N}
{\tens \mu}^{cc})$, where the first operator scales like
$T^\lambda \sim e^{-x\lambda}$ with a dimension
$\lambda= \lambda_{\rm sl}=N/f$, while the second is marginal
with $\lambda=0$. The detailed analysis of the other two
equations is much more complicated, but  one can still find their
exact solutions due to the simple structure of the $\L^{ab}$'s
Here we only briefly discuss the results of this
analysis.

It turns out that Eqs.~(\ref{eq:roscaling}) and (\ref{eq:tscaling})
have an infinite number of zero modes that can be divided into
two distinct classes. The first type corresponds to potential
scattering off the NLS and can be written as
\begin{equation}
 \dv^{ab}_{\rm pot} = \delta^{ab} \dv \;,
\label{eq:potscatt}
\end{equation}
where $\dv$ denotes an  arbitrary $\infty \times \infty$
Hermitian matrix. The rest of the  zero-modes can be
identified with the generators $\dv^{ab}_{\rm gen}$ of the 
unitary transformations of the $SU(N)$ Lie algebra,
Eq.~(\ref{eq:Lie}),
\begin{equation}
\dv^{ab}_{\rm gen} =
\left(\begin{array}{cc}
{\tens \varrho}^{ab}_{\rm gen} & \t^{ab}_{\rm gen} \\
(\t^{ba})^+_{\rm gen} & 0
\end{array}\right)
\;. \label{eq:generators}
\end{equation} 
More precisely, the generators ${\tens \varrho}^{ab}_{\rm gen}$
and $\t^{ab}_{\rm gen}$ 
can be shown to satisfy in first order the equations:
\begin{eqnarray}
&&\L^{ab}\t_{\rm gen}^{cd} - \L^{cd}\t_{\rm gen}^{ab} = \delta^{ad} \t_{\rm 
gen}^{cb} - 
\delta^{cb}\t_{\rm gen}^{ad} \;,\nonumber \\
&&[\L^{ab}, {\tens\varrho}_{\rm gen}^{cd}] +
[ \L^{cd}, {\tens\varrho}_{\rm gen}^{ab}] = \delta^{ad}
{\tens\varrho}_{\rm gen}^{cb} - \delta^{cb} {\tens\varrho}_{\rm
gen}^{ad} \;,
\end{eqnarray}
from which it follows that the operators ${\tilde \L}^{ab} := 
\L^{ab} + \dv^{ab}_{\rm gen}$ satisfy the same Lie algebra
as the original $\L^{ab}$'s:
\begin{equation} 
[{\tilde \L}^{ab},{\tilde \L}^{cd}] = \delta^{ad} {\tilde
\L}^{cb} - \delta^{cb} {\tilde \L}^{ad}\;. 
\end{equation} 
The ${\tens\varrho}_{\rm gen}^{ab}$'s turn out to be  the
generators of the unitary transformations in the $N$-dimensional
electronic subspace where the Lie-algebra Eq.~(\ref{eq:Lie}) is
realized, while the $\t^{ab}$'s correspond to the rotations of
this $N$-dimensional subspace.  

All the other eigenoperators around the fixed point can be shown
to be irrelevant. Very surprisingly, at least in the large $f$
limit, the leading irrelevant operators are quite different from
the leading irrelevant operator of the $SU(N)\times SU(f)$ model 
(\ref{eq:H_eff}), both in their structure and in their scaling
dimension. They are living in the sector $\t^{ab}$ and they can
be written as
\begin{equation}
{\dv}^{ab}_{\rm \;l} = \left(\begin{array}{cc}
0 & {\tens C}^{ab} \\
({\tens C}^{ba})^+ & 0
\end{array}\right)\;,
\label{eq:leading}
\end{equation}
where the $C^{ab}_{cm}$'s satisfy $\sum_a C^{aa}_{cm}=0$ and
$\sum_b (C^{ab}_{cm} - C^{cb}_{am})=0$ with $a,b,c=1,..,N$ and
$n=N+1,N+2, ..,\infty$. These operators have a dimension
\begin{equation}
\lambda_{\rm l}={N-1\over f} + \vartheta\left({N^2\over f^2}\right)
\; , 
\label{eq:lambda_l}
\end{equation}

We remark at this point that the operators (\ref{eq:leading}) do
{\it not exist} in the two-level-system model, which is
therefore completely equivalent to the corresponding
$SU(2)\times SU(f)$ model.\cite{ZarVlad} As we shall see, these
operators do not give a contribution to physical quantities
like the resistivity or the impurity susceptibility
but they influence the thermodynamic behavior of the model. 
We stress at this point, that their existence is strictly proven
in the $f\to \infty$ limit. They are very probably present even
in the $N<f$ case but  it is an open question if they survive
in the $N \le f$ limit. 

The impurity resistivity will be shown to be dominated by the
subleading  operators 
\begin{equation}
{\dv}_{\rm sl}^{ab} 
\sim \left(\begin{array}{cc}
{\tens Q}^{ab}& 0 \\
0 & {\tens S}^{ab}
\end{array}\right)\;,
\end{equation}
where the matrices ${\tens Q}^{ab}$ and ${\tens S}^{ab}$ satisfy
${\tens Q}^{aa}= {\tens S}^{aa}=0$ and $Q^{ab}_{dc}
=Q^{ab}_{dc}$. These operators have a dimension
\begin{equation}
\lambda_{\rm sl}={N \over f} + \vartheta\left({N^2\over
f^2}\right) \; .
\label{eq:lambda_sl}
\end{equation}
and the operator (\ref{eq:H_eff}) considered in the $SU(N)\times
SU(f)$ model is also one of them. Furthermore, one has other
even more irrelevant operators in the 
$\t$ sector of $\dv$ with a dimension $\lambda_{\rm
ssl}=(N+1)/f$ which give a subleading contribution to the physical
quantities calculated. 

In the previous considerations we did not take into account the
presence of the splitting $\Delta^{ab}$ of the NLS. As we
discussed already, this splitting results in the appearance of 
another low-energy scale, $T^*$. Below this the NLS cannot jump
freely between its $N$ different positions. Since usually the
ground state of the $NLS$ is non-degenerate in most cases 
a Fermi liquid state develops. In other words, the non-Fermi
liquid $SU(N)\times SU(f)$ fixed point is unstable with respect
to the splitting that usually drives the  system towards 
a Fermi liquid ground state. 

It has been argued very recently\cite{MF} that in
special cases, due to some dynamical Jahn-Teller effect, e.g., the
hopping amplitude $\Delta^{ab}$ might pick up and additional 
Berry phase, which could then result in a degenerate ground
state with degeneracy $N^\prime$. Then the effective Hamiltonian
at very low temperatures would be, of course, an
$SU(N^\prime)\times SU(f)$ exchange model, and in the region $T\ll T^*$ all our
previous considerations hold with the replacement of $N$ by
$N^\prime$. Unfortunately, this Berry phase scenario is not very
probable and therefore we expect the non-Fermi liquid behavior can 
be only observed  in the restricted temperature (energy) range
$T^*<\max\{T,\omega\} < T_K$, i.e., when the freezing temperature
is small enough. 

\begin{figure} 
\epsfxsize=7cm 
\hskip0.5cm\epsfbox{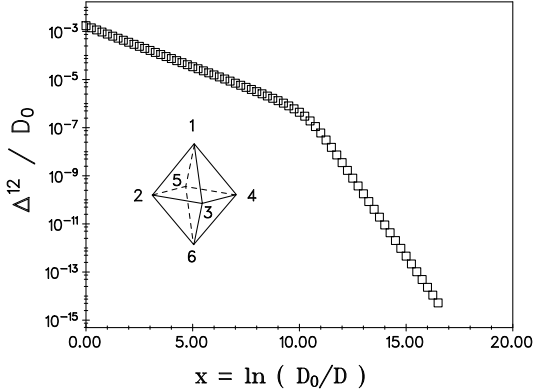}  
\vskip0.7truecm  
\caption{\label{fig:delta} Scaling of the  dimensionless  
hopping  
amplitude, $\Delta^{12}/D_0$ for the same 6-state system  
as in Fig.~\protect{\ref{fig:U's}}. Inset: Numbering of the  sites of the  
6-state system.} 
\end{figure}

Therefore it is very important to determine the realistic values of the
freezing temperature. We estimated the freezing temperature by
 solving the scaling equations (\ref{eq:scaling1}) and (\ref{eq:scaling2}) 
numerically for the same symmetrical 6-level system as in 
Fig.~\ref{fig:U's}. In this case the diagonal matrix elements
$\Delta^{aa}$ vanish by symmetry. As one can see from Fig.~\ref{fig:delta}, 
for a realistic NLS the renormalization of the 
hoppings $\Delta^{ab}$ is huge, and the situation 
$T^*<\max\{T,\omega\} < T_K$ can be reached  quite easily. 
We note at this point that in our Hamiltonian we also neglected
the contribution of two-electron scattering around the fixed
point, which might be also relevant in the immediate
neighborhood of the fixed point.\cite{MF}  However, these have a very
small amplitude and they are scaled downwards in the first part
of the scaling, $D>T_K$. Therefore, most likely their effect can be
neglected compared to that of the splitting $\Delta^{ab}$, which
provides the dominant mechanism  driving the system
to a Fermi liquid state.\cite{ZawNoz}

\subsection{Scaling of the physical quantities of the NLS model
in the large $f$ limit}
\label{ss:physicalq}

Now we turn to the calculation of the physical quantities. In
this Subsection we shall determine different thermodynamic
quantities and the conduction electrons' scattering rate,
$1/\tau$, which is directly proportional to the impurity
contribution to the electrical resistivity, $R_{\rm imp}(T)$.
\begin{figure} 
\epsfxsize=7cm  
\hskip0.5cm\epsfbox{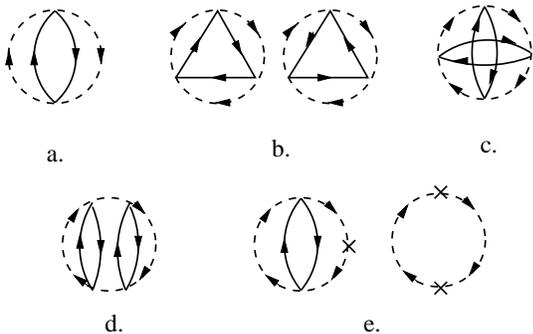}  
\vskip0.7truecm  
\caption{\label{fig:freeen} Diagrams generating the $1/f^2$ corrections to 
the free energy. The crosses denote the counterterm.} 
\end{figure}
To calculate a general physical quantity one should also
calculate the renormalization coefficient $A$ in
Eq.~(\ref{eq:mrg}), a non-trivial task away from
the fixed point. However, one can easily convince oneself that
the factors $A$ and $A^{-1}$ 
in the free energy corrections in Fig.~\ref{fig:freeen} and
the electronic self-energy corrections in Fig.~\ref{fig:nll}.c
  cancel exactly, and therefore these are
{\it scale invariant}, and can be calculated by considering only 
the scaling equations (\ref{eq:scaling1}) and
(\ref{eq:scaling2}). 

For the sake of simplicity let us assume first that the highest
low-energy scale is given by the temperature.
To calculate a physical quantity at a temperature  $T$
we apply a renormalization group transformation (\ref{eq:mrg})
with $D=D_0$ and $D^\prime = T$. Then in the new Hamiltonian 
all the logarithmic terms vanish since $\ln (D^\prime/T)=0$, and
the different physical quantities are exclusively given by the
{\it non-logarithmic contributions} of the corresponding
diagrams. For a scale invariant quantity like the free energy,
e.g., this implies that
\begin{eqnarray}
\lefteqn{F_{\rm imp}(D_0,T,\v_0^{ab},\Delta_0^{ab})=}&& \nonumber \\
&& F_{\rm imp} (T,T, \v^{ab}(\ln{D_0\over T}),
\Delta^{ab}(\ln{D_0\over T}))\;,
\end{eqnarray}
where on the right-hand-side no logarithmic corrections appear,
but the renormalized couplings have to be used. 

Therefore, in order to calculate the scaling behavior of the
thermodynamic quantities our task is to determine the
non-logarithmic  parts of the different free energy diagrams. 
Since the fixed point couplings $\v^{ab}_{\rm fp}$ are
proportional to $1/f$ up to $1/f^2$ order only the diagrams 
in Figs.~\ref{fig:freeen}.a-d. contribute.
However, these diagrams contain divergent contributions
originating from the finite part of the self-energy diagram 
in Fig.~\ref{fig:nll}.b. These spurious divergences can be
handled by a standard renormalization procedure,\cite{Itzykson} 
by adding the following  counterterm to the Hamiltonian:
\begin{equation}
H_{\rm count} = f\; 2D\;\ln2\; \chi^+_a \chi_c \tr\{\v^{ab}\v^{bc} \}
\;. 
\end{equation}
This counterterm can also be interpreted as a renormalization
of the bare parameters of the model, which should be used
in Eq.~(\ref{eq:mrg}) as the initial conditions. Then
the counterterm contributions in Fig.~\ref{fig:freeen}.e cancel
all the spurious divergences, and after a tedious calculation
one obtains for the non-logarithmic part of the free energy:
\begin{eqnarray}
F_{\rm imp}&=& -T \Bigl[ \ln N
+ {2\pi^2 f\over 3N}\bigl(\tr\{\v^{ab}\v^{bc}\v^{ca} -
\v^{ca}\v^{bc}\v^{ab}\}\bigr) \nonumber \\
&-& {f^2 \pi^2 \over 2N} 
\bigl(\tr\{\v^{ab}\v^{cd}\}\tr\{\v^{bc}\v^{da}\} \nonumber \\ &-&
\tr\{\v^{ab}\v^{bc}\}\tr\{\v^{cd}\v^{da}\}\bigr) + ...\Bigr]\;.
\label{eq:freeen}
\end{eqnarray}
Note that diagram (a) in Fig.~\ref{fig:freeen} is proportional
to $T^2/D$, and it does not give a contribution in the scaling
limit. 

Substituting the fixed point couplings Eq.~(\ref{eq:fp})
into Eq.~(\ref{eq:freeen}) the fixed point entropy can be
calculated as
\begin{eqnarray}
S_{\rm imp} &=& {\der F_{\rm imp} \over \der T} \approx
\ln N -{N^2-1 \over f^2} {\pi^2\over 6} 
\label{eq:s_imp}
\end{eqnarray}
which is just the expanded version of Eq.~(\ref{rsi2}). 
Note that Eq.~(\ref{eq:s_imp}) gives the NLS contribution to the
entropy only in the region $T^*\ll T\ll T_K$. Below 
$T^*$ the motion of the NLS is usually frozen out and the 
impurity entropy tends to zero, as expected for a
Fermi liquid state. 

The scaling of the free energy in the region $T^*\ll T\ll T_K$.
can be determined by expanding the $\v^{ab}$'s around their
fixed point values like Eq.~(\ref{eq:deviation}) and
substituting them into Eq.~(\ref{eq:freeen}). It turns out that
similarly to the multichannel Kondo and the two-level system
case\cite{Gan,ZarVlad} only the second order terms in
$\dv^{ab}$ contribute, and therefore in the temperature
range $T^*\ll T\ll T_K$ in leading order
the free energy and the specific heat scale as 
\begin{eqnarray}
&& F_{\rm imp}\sim T\left({T\over T_K}\right)^{2\lambda_{\rm
l}} \sim T^{{2(N-1)\over f}+1}\;, \label{eq:freeenscaling} \\
&& c_{\rm imp} \sim
\left({T\over T_K}\right)^{{2(N-1)\over f}}\;.
\end{eqnarray}
Below $T^*$ the Free energy generally shows a Fermi liquid behavior.
This scaling behavior does not agree with the one obtained in the 
BA solution of the exchange model. However, we have to remark 
at this point, that according to our estimations the 
amplitude of the subleading operators in $\dv^{ab}$ 
is {\it larger} than that of the leading irrelevant operators. Therefore,
one expects that there is a substantial energy region where
the subleading operators dominate, and eventually it is also possible
that they dominate the scaling of the  free energy in the whole 
region  $T^*\ll T\ll T_K$. 
Then the exponent 
$\lambda_{\rm l}$ in Eq.~(\ref{eq:freeenscaling}) should be
replaced by $\lambda_{\rm sl}$ and one obtains a scaling
$c_{\rm imp}\sim T^{2N/f}$ which is in $1/f$ order
completely identical with the Bethe ansatz and conformal field theory
results for the $SU(N)\times SU(f)$ model.

One can also easily determine the scaling of the splitting
susceptibility $\chi_\Delta = {\der^2 F_{\rm imp} /\; \der
\Delta^2}$ at $T=0$ for small $\Delta$'s, where now $\Delta$
denotes the characteristic value of the splittings
$\Delta^{ab}$. Investigation of the free energy diagrams
Fig.~\ref{fig:freeen} shows that the 'splitting  magnetization'
$M_\Delta ={\der F_{\rm imp} /\; \der \Delta}$ should be of
the form:
\begin{equation}
M_\Delta = m({\Delta\over D}, \v^{ab})\;.
\label{eq:m}
\end{equation}
The important point is that  $\Delta$ is not scale
invariant, but it rather behaves as
\begin{equation}
\Delta^\prime = Z_{\Delta}\left({D_0\over D}, \v^{ab}\right) \;
\Delta\;, 
\end{equation}
where the factor $Z_\Delta$ should be determined by integrating
Eq.~(\ref{eq:scaling1}). As a consequence, $M_\Delta$ is {\it not
scale invariant} either, and it has to be rescaled under the RG
transformation by the factor $Z_\Delta$. Therefore, applying the
renormalization group transformation to Eq.~(\ref{eq:m}) with 
$D^\prime=\Delta^*=T^*$ we obtain:
\begin{equation}
M_\Delta= Z_{\Delta}\left({D_0\over \Delta^* }, \v^{ab}\right)
\times m(1,\v^{ab}_{\rm fp})\;,
\end{equation}
where we assumed that $\Delta^*\ll T_K$ and thus the scaled couplings 
$\v^{ab}(D^\prime)$ can be replaced by their fixed point values.
Since $m(1,\v^{ab}_{\rm fp})$ is just a constant, the
scaling of $M_\Delta$ is the same as that of the factor 
$Z_{\Delta}\left({D_0\over \Delta^* }, \v^{ab}\right)$. For
very small $\Delta$'s the scaling of $Z_\Delta$ can be easily 
determined from the fixed point form of the scaling
equation~(\ref{eq:scaling1}) 
\begin{equation}
{d\Delta^{ab}\over dx} = - {N\over f} \Delta^{ab}\;,
\end{equation}
and one obtains in leading order in $1/f$:
\begin{equation}
M_\Delta \sim Z_\Delta({D_0\over \Delta^*})\sim \left(
{\Delta^*\over T_K}\right)^{N/f}\approx 
\left({\Delta \over T_K}\right)^{N/f}\;,
\label{eq:M}
\end{equation}
in agreement with (\ref{cv}) and the conformal field
theory results.\cite{AL} In higher order in $1/f$ one also has
to take into account the renormalization of the splitting
in Eq.~(\ref{eq:M}), $\Delta^*\sim \Delta^{1/(1-\lambda_{\rm
sl})}$ and one obtains with $\lambda_{\rm sl}={N\over
N+f}\approx {N\over f} - {N^2\over f^2}$ 
\begin{equation}
M_{\Delta}\sim \Delta ^{{\lambda_{\rm sl}\over 1- \lambda_{\rm
sl}}} \sim \Delta^{N/f}\;,
\end{equation}
which is the exact result.\cite{Cox,AL}

Finally, we discuss the scaling of the electronic scattering
rate, which we determine from the imaginary part of the electronic
self-energy in Fig.~\ref{fig:nll}.c. By assuming a finite impurity
concentration $n_i$ and averaging over the position of the impurities
and the orientation of the incoming electrons we obtain for the
average scattering rate:
\begin{equation}
\langle {1\over \tau} \rangle = 2\pi n_i (2D_0 )
{1\over N}\tr\{\v^{ab}\v^{ba}\}\;.
\end{equation}
Note, that the factor $D_0$ arises from the inverse density of states 
$\varrho_0^{-1}$ and is invariant under scaling. Substituting
into this equation $\v^{ab}=\v^{ab}_{\rm fp} + \dv^{ab}$
we see immediately that the leading irrelevant operators do not
give a contribution to the electronic scattering rate which
is dominated by  subleading operators and scales like
\begin{eqnarray}
{1\over \tau } & \sim & T^{\lambda_{\rm sl}} \sim
T^{N/f}\phantom{nnn}  (\omega=0)\;, \nonumber \\
{1\over \tau } & \sim & \omega^{N/f}\phantom{nnn} (T=0)\;.
\end{eqnarray}
In higher orders this result should be replaced by $1/\tau \sim
T^{N/(f +N)}$ and $1/\tau \sim \omega^{N/(f +N)}$. 

\subsection{Discussion of the possible physical realizations of the
NLS model}
\label{ss:physreal}

The simplest possible realization of the $SU(N)\times SU(f)$ model
is given by substitutional impurities in metals. These
impurities may form tunneling centers\cite{NLS,Fukuyama} which
then interact  with the conduction electrons' band. 
An example of such a system is given by $Pb_{1-x} Ge_x
Te$.\cite{Fukuyama} 
The alloy $PbTe$ is a narrow gap semiconductor, but usually
because of some intrinsic impurities it becomes metallic at low
temperatures. Since the $Ge^{2+}$ ions are smaller than the
$Pb^{2+}$ ions and they are also attracted by their nearest
neighbor $Te^{2-}$ ions, they form 8-state systems, and according to the
our discussions in Section~\ref{ss:fixpoint} they
would be good candidates for the $SU(8)\times SU(2)$ model. 

However, while an unambiguous logarithmic anomaly has been
observed in the resistivity of these materials,\cite{Fukuyama} 
no non-Fermi liquid behavior has been detected. There may be several 
reasons for that. According to  the results of the BA calculations in the 
case $N>f$ the non-Fermi-liquid  corrections are {\it subleading}, and the 
physical quantities have  in leading order a Fermi liquid-like behavior. 
The subleading low-temperature behavior of these alloys has never been 
analyzed and the original measurements do not seem to be accurate enough to 
extract such detailed information. We are also not aware of any 
measurements of other physical quantities like the specific heat in the 
interesting concentration domain.  Furthermore, $PbTe$ has 
very complicated properties: it has a soft phonon mode that drives the 
system trough a ferroelectric phase transition as a function of the $Ge$
concentration, and  strong spin-orbit scattering 
which most likely spoils the $SU(2)$ symmetry of the electron
spins as well. Moreover, the measurements have been carried out at
relatively large $Ge$ concentrations, where the interaction of
the NLS's can not be neglected any longer. 

It seems  that in order to observe the non-Fermi liquid
scaling much more accurate measurements should be carried out 
at even lower temperatures and {\it lower} $Ge$ concentrations
for {\it several} physical quantities
and {\it lower} $Ge$ concentrations. One could also try to find
a better candidate. Since in the case of $PbTe$ the formation of the 
NLS's is induced by the ionic attractions, we think that experimentalists 
should search  among multicomponent metals, where  interstitials can
occur in the material. 
\begin{figure} 
\epsfxsize=7cm  
\hskip0.5cm\epsfbox{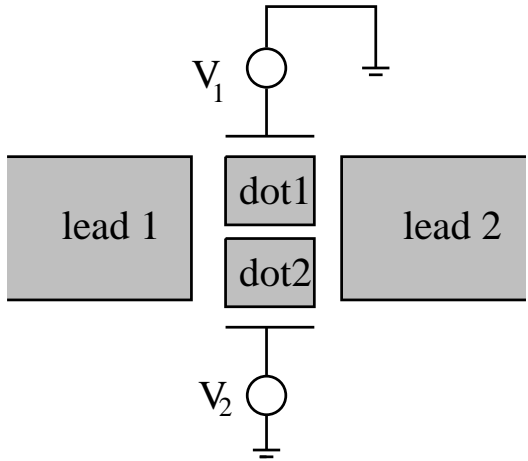}  
\vskip0.7truecm  
\caption{\label{fig:dots} 
A mesoscopic double dot system, candidate for the $SU(3)\times SU(2)$ model.} 
\end{figure}
Similarly to the case of the two-channel Kondo model,\cite{Matveev} another
possible realization of the $SU(N)\times SU(f)$ model could be possible by 
means of nanotechnologies. In Fig.~\ref{fig:dots} we show a 
double dot geometry which is a candidate for the realization of the 
$SU(3)\times SU(2)$ model. In the leads the electrons can be
described as free particles: 
\begin{equation}
H_{\rm leads} = \sum_{\alpha=l1,l2} \sum_{\epsilon \sigma} \epsilon \;
c^+_{\epsilon \alpha \sigma} c_{\epsilon   \alpha \sigma}\;,
\end{equation}
where $\alpha=l1,l2$ refers to the two leads and $\sigma=\pm$ is
the electron spin. 

The Hamiltonian of the dots can be written as\cite{Devoret} 
\begin{eqnarray}
H_{\rm dots}&=&  \sum_{\alpha=d1,d2} \sum_{\epsilon \sigma} \epsilon \;
c^+_{\epsilon \alpha \sigma} c_{\epsilon   \alpha \sigma}\;\nonumber\\
& + & {(Q_1-V_1 C_1)^2\over 2C_{\Sigma,1}} + {(Q_2-V_2 C_2)^2\over
2C_{\Sigma,2}} + {Q_1Q_2\over C_{12}}\;,
\end{eqnarray}
where $\alpha=d1,d2$ is the index of the two dots, $Q_1$ and
$Q_2$ are the charges of them, $V_1$ and $V_2$ denote the
applied gate voltages in the Figure, and the $C$'s denote different
capacitances of the system.\cite{mesoscopic} With a suitable choice of 
the gate voltages one can achieve that the ground state of the
dots becomes three fold degenerate corresponding to the states 
$\beta:=(Q_1,Q_2)=(0,0)$, $(0,e)$ and $(e,0)$. 
Then the tunneling processes among the leads and the dots result
in the simultaneous flips  in the electrons orbital quantum
number $\alpha=\{l1,l2,d1,d2 \}$ and the charge variables $\beta$ which
now take over the role of the orbital index of the NLS. Since
the tunneling is diagonal in the electron's spin we have an
additional $SU(2)$ degeneracy in the spin of the electrons. 
In sum, this system is a good candidate for the realization of
the $SU(3)\times SU(2)$ model, where the $SU(3)$ fixed point
symmetry corresponds to the three fold degenerate ground states
of the dots.  

\section{Conclusions} 
We have studied the Multichannel Coqblin-Schrieffer model (MCCS) and
its relation to the N-level system model (NLS). The properties of
the MCCS model depend on both the spin and flavor symmetries, $SU(N)$
and $SU(f)$, as well as on the {\em spin} of the impurity. 

We have performed both analytical and numerical studies of the model.
As with the
multichannel Kondo model, there are three different classes of fixed
points depending on the spin of the impurity, $\mu$. The underscreened
and completely screened fixed points ($\mu > f$ and $\mu = f$,
respectively) have qualitative similar behavior to the analogous
multichannel counterparts ($N=2$). 

There are overscreened fixed points. They display
Non-Fermi liquid behavior: they have associated  anomalous residual
entropy and 
anomalous exponents in the  low-temperature expansion of quantities
like the specific heat and the magnetic susceptibility. 

For an
impurity with spin in the fundamental representation of $SU(N)$, the
residual entropy, ${\cal{S}}^i$, is only a function of $N+f$ and $|\log
(N/f)|$. Hence, there are different fixed points with the same value
of ${\cal{S}}^i$. The exception corresponds to $N=f$, which yields the
largest value of the residual entropy for fixed $N+f$.  

The low-temperature thermodynamics are determined by the value of the
ratio $\gamma=f/N$ alone, for any $\mu < f$. When $f \ne N$ we have
\begin{eqnarray}
\frac{C_V^i}{T} ,~ \chi^i \propto ctn.+ \left( \frac{T}{T_K}
\right)^{\frac{1-\gamma}{1+\gamma}}
\label{fin1} 
\end{eqnarray}
which diverge for $\gamma > 1$, but remain finite for $\gamma <
1$. When $N = f$, the power is replaced by a logarithm, as in the
two-channel Kondo model. The constant terms in (\ref{fin1}) are always
present and  are the dominant contributions in the
completely screened case, $\mu = f$, and when $f < N$. The Wilson
ratio in such cases is given by
\begin{eqnarray*}
R = \frac{N(N+f)}{N^2-1}
\end{eqnarray*}

Channel anisotropy is a relevant perturbation. As the channel symmetry
is reduced from $SU(f)$ to $SU(f')$, the entropy is quenched since
$N+f$ decreases and $|\log (N/f)|$ increases. Likewise, a system with
channel anisotropy might behave like a $f>N$ system at intermediate
temperatures and flow at low-$T$ to a $f<N$ system.


Then we turned to the comparison of the MCCS model to the N-level system 
model (NLS), describing a heavy particle tunneling between $N$ different 
positions and interacting with the conduction electrons. We have shown that
the low-energy fixed point of the NLS model is just the $SU(N)\times SU(f)$
MCCS model.  Performing a $1/f$ study of the NLS model 
we have analyzed the operator content of this low-energy fixed point,
and the scaling properties of different physical quantities in the
$N<f$ limit. We have shown in this limit that while the operator 
content of the NLS model is different from that of the MCCS model, 
apart from some subtle differences, the low-energy properties of the 
two models are the same.  Especially, comparison with the exact results 
obtained in the first part of the paper and with the NCA 
calculations\cite{Cox} show that the susceptibility, the residual entropy 
and the resistivity of the two models behave in the same 
way, and for reasonable physical parameters even the scaling of the specific 
heat is properly described by the MCCS model.

Finally, we discussed some  possible physical realizations of the $SU(N)
\times SU(f)$ models. First we discussed the case of tunneling 
interstitials in multicomponent metals such as $Pb_{1-x}Ge_xTe$ 
compounds. We pointed out that the low concentration of interstitials 
is essential to avoid strong inter-impurity interactions and keep
the diagonal elements of the self-energy $\Delta^{ab}$ small. 
Secondly, we suggested a double quantum dot structure that could give
an ideal realization for the $SU(3)\times SU(2)$ model.

\section{Acknowledgments}

In the course of this investigation we learned of parallel work by
A. Georges, O. Parcollet, G. Kotliar and A. Sengupta, using conformal
field theory and large-N approach to study the same model.
There is perfect agreement whenever comparison can be made. 
We are most grateful to the authors for many useful and 
enlightening discussions and sharing their results prior to circulation. 
Part of the work was carried out while N.A. was visiting the group of 
Physique Theorique at the ENS. It is a pleasure to thank  the members
of the group for their warm hospitality. 

G.Z. would like to acknowledge useful discussions  
with D.L. Cox  K. Vlad\'ar, A. Zawadowski, and A. Moustakas.
He would like to thank the Magyary Zolt\'an Foundation 
and the Institut  Laue-Langevin (Grenoble)  for its hospitality,
where part of the present work  has been done. 

A.J. would like to acknowledge useful discussions with R. Bulla, 
P. Coleman, F.H.L. Essler, A. Hewson, A.F. Ho, A. Lopez, 
P. Nozi\`{e}res, R. Ramazashvili, 
and A. Tsvelik. The numerical
calculations have been carried out using the computing facilities 
of Theoretical Physics, University of Oxford.

N.A. is grateful the C. Destri and D. Braak for sharing their insights
in the course many illuminating 
discussions, and to A. Ruckenstein for a careful reading of the manuscript.

The research has been supported by the EPSRC grant GR/K97783, the 
Hungarian Grants OTKA F016604 and 
OTKA 7283/93, and the U.S-Hungarian Joint Fund No.587.

\end{document}